\documentclass[twocolumn,english]{IEEEtran}
\usepackage[T1]{fontenc}
\usepackage[latin9]{inputenc}
\usepackage{array}
\usepackage{rotating}
\usepackage{multirow}
\usepackage{amssymb}
\usepackage{epstopdf}
\usepackage{graphicx}

\makeatletter

\newcommand{\noun}[1]{\textsc{#1}}
\providecommand{\tabularnewline}{\\}
\@ifundefined{lettrine}{\usepackage{lettrine}}{}

\@ifundefined{showcaptionsetup}{}{%
 \PassOptionsToPackage{caption=false}{subfig}}
\usepackage{subfig}
\makeatother

\begin{document}

\title{Machine Learning in Wireless Sensor Networks: Algorithms, Strategies, and Applications}

\author{Mohammad Abu Alsheikh\textsuperscript{1,2}, Shaowei Lin\textsuperscript{2}, Dusit Niyato\textsuperscript{1} and Hwee-Pink Tan\textsuperscript{2}}
\IEEEspecialpapernotice{\textsuperscript{1}School of Computer Engineering, Nanyang Technological University, Singapore 639798\\
\textsuperscript{2}Sense and Sense-abilities Programme, Institute for Infocomm Research, Singapore 138632}
\maketitle

\begin{abstract}
Wireless sensor networks monitor dynamic environments that change rapidly over time. This dynamic behavior is either caused by external factors or initiated by the system designers themselves. To adapt to such conditions, sensor networks often adopt machine learning techniques to eliminate the need for unnecessary redesign. Machine learning also inspires many practical solutions that maximize resource utilization and prolong the lifespan of the network. In this paper, we present an extensive literature review over the period 2002-2013 of machine learning methods that were used to address common issues in wireless sensor networks (WSNs). The advantages and disadvantages of each proposed algorithm are evaluated against the corresponding problem. We also provide a comparative guide to aid WSN designers in developing suitable machine learning solutions for their specific application challenges.
\end{abstract}

\begin{IEEEkeywords}
Wireless sensor networks, machine learning, data mining, security, localization, clustering, data aggregation, event detection, query processing, data integrity, fault detection, medium access control, compressive sensing.
\end{IEEEkeywords}

\section{INTRODUCTION}

\lettrine{A} wireless sensor network (WSN) is composed typically of multiple autonomous, tiny, low cost and low power sensor nodes. These nodes gather data about their environment and collaborate to forward sensed data to centralized backend units called base stations or sinks for further processing. The sensor nodes could be equipped with various types of sensors, such as thermal, acoustic, chemical, pressure, weather, and optical sensors. Because of this diversity, WSNs have tremendous potential for building powerful applications, each with its own individual characteristics and requirements. Developing efficient algorithms that are suitable for many different application scenarios is a challenging task. In particular, WSN designers have to address common issues related to data aggregation, data reliability, localization, node clustering, energy aware routing, events scheduling, fault detection and security.

Machine learning (ML) was introduced in the late 1950\textquoteright{}s as a technique for artificial intelligence (AI) \cite{ayodeleintroduction2010}. Over time, its focus evolved and shifted more to algorithms which are computationally viable and robust. In the last decade, machine learning techniques have been used extensively for a wide range of tasks including classification, regression and density estimation in a variety of application areas such as bioinformatics, speech recognition, spam detection, computer vision, fraud detection and advertising networks. The algorithms and techniques used come from many diverse fields including statistics, mathematics, neuroscience, and computer science. The following two classical definitions capture the essence of machine learning:
\begin{enumerate}
\item The development of computer models for learning processes that provide solutions to the problem of knowledge acquisition and enhance the performance of developed systems \cite{590079}.
\item The adoption of computational methods for improving machine performance by detecting and describing consistencies and patterns in training data \cite{Langley:1995:AML:219717.219768}. 
\end{enumerate}
Applying these definitions to WSNs, we see that the promise of machine learning lies in exploiting historical data to improve
the performance of sensor networks on given tasks without the need for re-programming. More specifically, machine learning is important in WSN applications 
for the following main reasons:
\begin{enumerate}
	\item Sensor networks usually monitor dynamic environments that change rapidly over time. For example, a node's location may change due to soil erosion or sea turbulence. It is desirable to develop sensor networks that can adapt and operate efficiently in such environments.
	\item WSNs may be used for collecting new knowledge about unreachable, dangerous locations \cite{451250} (e.g., volcano eruption and waste water monitoring) in exploratory applications. Due to the unexpected behavior patterns that may arise in such scenarios, system designers may develop solutions that initially may not operate as expected. System designers would rather have robust machine learning algorithms that are able to calibrate itself to newly acquired knowledge.
\item WSNs are usually deployed in complicated environments where researchers cannot build accurate mathematical models to describe the system behavior. Meanwhile, some tasks in WSNs can be prescribed using simple mathematical models but may still need complex algorithms to solve them (e.g., the routing problem \cite{1030829,1368893}). Under similar circumstances, machine learning provides low-complexity estimates for the system model.	
\item Sensor network designers often have access to large amounts of data but may be unable to extract important correlations in them. For example, in addition to ensuring communication connectivity and energy sustainability, the WSN application often comes with minimum data coverage requirements that have to be fulfilled by limited sensor hardware resources \cite{romer2004design}. Machine learning methods can then be used to discover important correlations in the sensor data and propose improved sensor deployment for maximum data coverage.
	\item New uses and integrations of WSNs, such as in cyber-physical systems (CPS), machine-to-machine (M2M) communications, and Internet of things (IoT) technologies, have been introduced with a motivation of supporting more intelligent decision-making and autonomous control \cite{wan2013machine}. Here, machine learning is important to extract the different levels of abstractions needed to perform the AI tasks with limited human intervention \cite{bengio2009learning}.
\end{enumerate}

However, there are a few drawbacks and limitations that should be considered when using machine learning techniques in wireless sensor networks. Some of these are:
\begin{enumerate}
	\item As a resource limited framework, WSN drains a considerable percentage of its energy budget to predict the accurate hypothesis and extract the consensus relationship among data samples. Thus, the designers should consider the trade-off between the algorithm's computational requirements and the learned model's accuracy. Specifically, the higher the required accuracy, the higher the computational requirements, and the higher energy consumptions. Otherwise, the developed systems might be employed with centralized and resource capable computational units to perform the learning task.
	\item Generally speaking, learning by examples requires a large data set of samples to achieve the intended generalization capabilities (i.e., fairly small error bounds), and the algorithm's designer will not have the full control over the knowledge formulation process \cite{hoffmann1990general}.
\end{enumerate}

During the past decade, WSNs have seen increasingly intensive adoption of advanced machine learning techniques. In \cite{4449882}, a short survey of machine learning algorithms applied in WSNs for information processing and for improving network performance was presented. A related survey that discussed the applications of machine learning in wireless ad-hoc networks was published in \cite{4496871}. The authors of \cite{forster2010machine} discussed applications of three popular machine learning algorithms (i.e., reinforcement learning, neural networks and decision trees) at all communication layers in the WSNs. In contrast, specialized surveys that touch on machine learning usage in specific WSN challenges have also been written. For instance, \cite{5451757,4589621} addressed the development of efficient outlier detection techniques so that proper actions can be taken, and some of these techniques are based on concepts from machine learning. Meanwhile, \cite{5473889} discusses computational intelligence methods for tackling challenges in WSNs such as data aggregation and fusion, routing, task scheduling, optimal deployment and localization. Here, computational intelligence is a branch of machine learning that focuses on biologically-inspired approaches such as neural networks, fuzzy
systems and evolutionary algorithms \cite{74851240}.

Generally, these early surveys concentrated on reinforcement learning, neural networks and decision trees which were popular due to their efficiency in both theory and practice. In this paper, we decided instead to include a wide variety of important \textit{up-to-date} machine learning algorithms for a comparison of their strengths and weaknesses. In particular, we provide a  comprehensive overview which groups these recent techniques roughly into supervised, unsupervised and reinforcement learning methods. Another distinction between our survey and earlier works is the way that machine learning techniques are presented. Our work discusses machine learning algorithms based on their target WSN challenges, so as to encourage the adoption of existing machine learning solutions in WSN applications. Lastly, we build on existing surveys and go beyond classifying and comparing previous efforts, by providing useful and practical guidelines for WSN researchers and engineers who are interested in exploring new machine learning paradigms for future research. 

The rest of the paper is organized as follows: 
\begin{itemize}
\item Section \ref{sec:overview_ml} introduces the reader to machine learning algorithms and themes that will be referred to in later sections. Simple examples will be given in the context of WSNs.
\item In Section \ref{sec:functional_challenges}, we review existing machine learning efforts to address \textit{functional} issues in WSNs such as routing, localization, clustering, data aggregation, query processing and medium access control. Here, an issue is functional if it is essential to the basic operation of the wireless sensor network.
\item Section \ref{sec:non_functional_challenges} investigates machine learning solutions in WSNs for fulfilling \textit{non-functional} requirements, i.e. those which determine the quality or enhance the performance of functional behaviors. Examples of such requirements include security, quality of service (QoS)
and data integrity. In this section, we also highlight some unique efforts in specialized WSN applications.
\item Section \ref{sec:open_issues} outlines major difficulties and open research problems for machine learning in WSNs.
\item Finally, we conclude in Section \ref{sec:CONCLUSIONS-AND-FUTURE} and present a comparative guide with useful paradigms for furthering machine learning research in various WSN applications.
\end{itemize}

\section{\label{sec:overview_ml}INTRODUCTION TO MACHINE LEARNING IN WIRELESS SENSOR NETWORKS}

Usually, sensor network designers characterize machine learning as a collection of tools and algorithms that are used to create prediction models. However, machine learning experts recognize it as a rich field with very large themes and patterns. Understanding such themes will be beneficial to those who wish to apply machine learning to WSNs. Applied to numerous WSNs applications, machine learning algorithms provide tremendous flexibility benefits. This section provides some of the theoretical concepts and strategies of adopting machine learning in the context of WSNs.

Existing machine learning algorithms can be categorized by the intended structure of the model. Most machine learning algorithms fall into the categories of supervised, unsupervised and reinforcement learning \cite{Abu-Mostafa:2012:LD:2207825}. In the first category, machine learning algorithms are provided with a labeled training data set. This set is used to build the system model representing the learned relation between the input, output and system parameters. In contrast to supervised learning, unsupervised learning algorithms are not provided with labels (i.e., there is no output vector). Basically, the goal of an unsupervised learning algorithm is to classify the sample sets to different groups (i.e., clusters) by investigating the similarity between the input samples. The third category includes reinforcement learning algorithms, in which the agent learns by interacting with its environment (i.e., online learning). Finally, some machine learning algorithms do not naturally fit into this classification since they share characteristics of both supervised and unsupervised learning methods. These hybrid algorithms (often termed as semi-supervised learning) aim to inherit the strengths of these main categories, while minimizing their weaknesses \cite{chapelle2006semi}.

This section is mainly to introduce the reader to the algorithms that will be referred to in later sections. Moreover, examples will be given to demonstrate the process of adopting machine learning in WSNs. In Sections \ref{sec:functional_challenges} and \ref{sec:non_functional_challenges}, such details will be omitted. For interested reader, please refer to \cite{Abu-Mostafa:2012:LD:2207825,720536} and references therein, for thorough discussions of machine learning theory and its classical concepts.

\subsection{Supervised Learning}

In supervised learning, a labeled training set (i.e., predefined inputs and known outputs) is used to build the system model. This model is used to represent the learned relation between the input, output and system parameters. In this subsection, the major supervised learning algorithms are discussed in the context of WSNs. In fact, supervised learning algorithms are extensively used to solve several challenges in WSNs such as localization and objects targeting (e.g., \cite{4518167,5161349,Shareef:2008:LUN:1361492.1361497}), event detection and query processing (e.g., \cite{1541008,5698542,1544272,5702151}), media access control (e.g., \cite{5612365,Shen2008900,Kulkarni:2009:NNB:1704555.1704768}), security and intrusion detection (e.g., \cite{1665221,1648838,4496866,4289308}), and quality of service (QoS), data integrity and fault detection (e.g., \cite{1606090,4429838,Wang:2007:PLQ:1317425.1317434}).

\subsubsection{K-nearest neighbor (k-NN)}

This supervised learning algorithm classifies a data sample (called a query point) based on the labels (i.e., the output values) of the near data samples. For example, missing readings of a sensor node can be predicted using the average measurements of neighboring sensors within specific diameter limits. There are several functions to determine the nearest set of nodes. A simple method is to use the Euclidean distance between different sensors. K-nearest neighbor does not need high computational power, as the function is computed relative to local points (i.e., k-nearest points, where k is a small positive integer). This factor coupled with the correlated readings of neighboring nodes makes k-nearest neighbor a suitable distributed learning algorithm for WSNs. In \cite{beyer1999nearest}, it has been shown that the k-NN algorithm may provide inaccurate results when analyzing problems with high-dimensional spaces (more than 10-15 dimensions) as the distance to different data samples becomes invariant (i.e., the distances to the nearest and farthest neighbors are slightly similar). In WSNs, the most important application of the k-nearest neighbor algorithm is in the query processing subsystem (e.g., \cite{1541008,5698542}). 

\subsubsection{Decision tree (DT)}

It is a classification method for predicting labels of data by iterating the input data through a learning tree \cite{ayodele2010types}. During this process, the feature properties are compared relative to decision conditions to reach a specific category. The literature is very rich with solutions that use DT algorithm to resolve different WSNs' design challenges. For example, DT provides a simple, but efficient method to identify link reliability in WSNs by identifying a few critical features such as loss rate, corruption rate, mean time to failure (MTTF) and mean time to restore (MTTR). However, DT works only with linearly separable
data and the process of building optimal learning trees is NP-complete \cite{safavian1991survey}.

\subsubsection{Neural networks (NNs)}

This learning algorithm could be constructed by cascading chains of decision units (e.g., perceptrons or radial basis functions) used to recognize non-linear and complex functions \cite{bengio2009learning}. In WSNs, using neural networks in distributed manners is still not so pervasive due to the high computational requirements for learning the network weights, as well as the high management overhead. However, in centralized solutions, neural networks can learn multiple outputs and decision boundaries at once \cite{lippmann1987introduction}, which makes them suitable for solving several network challenges using the same model.

\begin{figure}[t]
\begin{centering}
\includegraphics[trim=1.5cm 0.8cm 1.5cm 0.4cm, width=0.75\columnwidth]{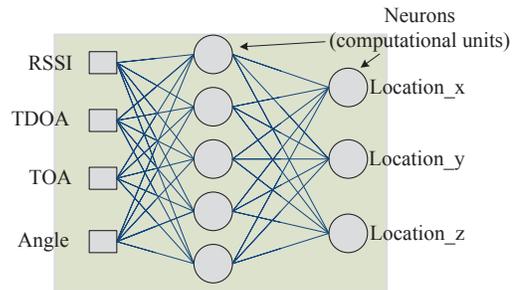}
\par\end{centering}
\caption{\label{fig:neural_network_example}Illustration example of node localization in WSNs in 3D space using supervised neural networks.}
\end{figure}

We consider a sensor node localization problem (i.e., determining node's geographical position) as an application example of neural network in WSNs. Node localization can be based on propagating angle and distance measurements of the received signals from anchor nodes \cite{8547966}. Such measurements may include received signal strength indicator (RSSI), time of arrival (TOA), and time difference of arrival (TDOA) as illustrated in Figure \ref{fig:neural_network_example}. After supervised training, neural network generates an estimated node location as vector-valued coordinates in 3D space. Related algorithms to neural networks include self-organizing map (or Kohonen's maps) and learning vector quantization (LVQ) (see \cite{Kohonen:2001} and references therein for an introduction to these methods). In addition to function estimation, one of the important applications of neural networks is for big data (high-dimensional and complex data set) tuning and dimensionality reduction \cite{hinton2006reducing}.

\subsubsection{Support vector machines (SVMs)}

\begin{figure}[t]
\begin{centering}
\includegraphics[trim=0.5 0.6cm 0.5cm 0.5cm, width=0.85\columnwidth]{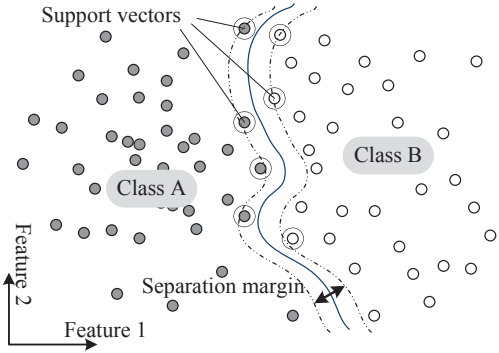}
\par\end{centering}
\caption{\label{fig:nonlinear_svm}An example of non-linear support vector machines.}
\end{figure}

It is a machine learning algorithm that learns to classify data points using labeled training samples \cite{steinwart2008support}. For example, one approach for detecting malicious behavior of a node is by using SVM to investigate temporal and spatial correlations of data. To illustrate, given WSN's observations as points in the feature space, SVM divides the space into parts. These parts are separated by as wide as possible margins (i.e., separation gaps), and new reading will be classified based on which side of the gaps they fall on as shown in Fig. \ref{fig:nonlinear_svm}. An  SVM algorithm, which includes optimizing a quadratic function with linear constraints (that is, the problem of constructing a set of hyperplanes), provides an alternative method to the multi-layer neural network with non-convex and unconstrained optimization problem \cite{ayodele2010types}. Potential applications of SVM in WSNs are security (e.g., \cite{4496866,4289308,4761978,2563258,Zhang20131062}) and localization (e.g., \cite{5506589,4384476,5487032}). For a detailed discussion of the SVM theory, please refer to \cite{steinwart2008support}.

\subsubsection{Bayesian statistics}

Unlike most machine learning algorithms, Bayesian inference requires a relatively small number of training samples \cite{box2011bayesian}. Bayesian methods adapt probability distribution to efficiently learn uncertain concepts (e.g., $\theta$) without over-fitting. The crux of the matter is to use the current knowledge (e.g., collected data abbreviated as $D$) to update prior beliefs into posterior beliefs 
$p(\theta|D)\propto p(\theta)p(D|\theta)$, where $p(\theta|D)$ is the posterior probability of the parameter $\theta$ given the observation $D$, and $p(D|\theta)$ is the likelihood of the observation $D$ given the parameter $\theta$. One application of Bayesian inference in WSNs is assessing event consistency ($\theta$) using incomplete data sets ($D$) by investigating prior knowledge about the environment. However, such statistical knowledge requirement limits the wide adoption of Bayesian algorithms in WSNs. A related statistical learning algorithm is Gaussian process regression (GPR) model~\cite{Rasmussen06gaussianprocesses}.

\subsection{Unsupervised Learning}

Unsupervised learners are not provided with labels (i.e., there is no output vector). Basically, the goal of an unsupervised learning algorithm is to classify the sample set into different groups by
investigating the similarity between them. As expected, this theme of learning algorithms is widely used in node clustering and data aggregation problems (e.g., \cite{5489603,5425458,5345599,5706781,5671089,4249813,985674}). Indeed, this wide adoption is due to data structures (i.e., no labeled data is available) and the desired outcome in such problems.

\subsubsection{K-means clustering}

The k-means algorithm \cite{kanungo2002efficient} is used to recognize data into different classes (known as clusters). This unsupervised learning algorithm is widely used in sensor node clustering problem due to its linear complexity and simple implementation. The k-means steps to resolve such node clustering problem are (a) randomly choose $k$ nodes to be the initial centroids for different clusters; (b) label each node with the closest centroid using a distance function; (c) re-compute the centroids using the current node memberships and (d) stop if the convergence condition is valid (e.g., a predefined threshold for the sum of distances between nodes and their perspective centroids), otherwise go back to step (b). 

\subsubsection{Principal component analysis (PCA)}

\begin{figure}[t]
\begin{centering}
\includegraphics[trim=1.5cm 0.8cm 1.5cm 0.4cm, width=0.7\columnwidth]{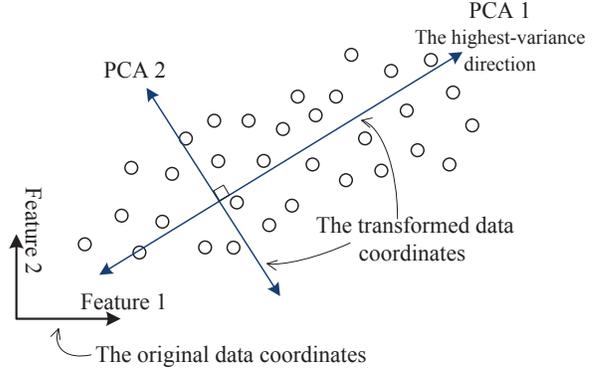}
\par\end{centering}
\caption{\label{fig:pca_example}A simple 2D visualization of the principal component analysis algorithm. It is important to note that the potential of the PCA algorithm is high mainly when dealing with high-dimensional data \cite{jolliffe2002principal}.}
\end{figure}

It is a multivariate method for data compression and dimensionality reduction that aims to extract important information from data and present it as a set of new orthogonal variables called principal components \cite{jolliffe2002principal}. As shown in Fig. \ref{fig:pca_example}, the principal components are ordered such that the first component corresponds to the highest-variance direction of the data, and so on for the other components. Hence, the least-variance components can be discarded as they contain the least information content. For example, PCA reduces the amount of transmitted data among sensor nodes by finding a small set of uncorrelated linear combinations of original readings. Furthermore, the PCA method simplifies the problem solving by considering only few conditions in very large variable problems (i.e., tuning big data into tiny data representation) \cite{feldman2013turning}. A thorough discussion of the PCA theory (e.g., the eigenvalue, eigenvector, and covariance matrix analysis) is given in \cite{jolliffe2002principal}.

\subsection{Reinforcement Learning}

\begin{figure}[t]
\begin{centering}
\includegraphics[trim=0.5 0.8cm 0.5cm 0.5cm, width=1\columnwidth]{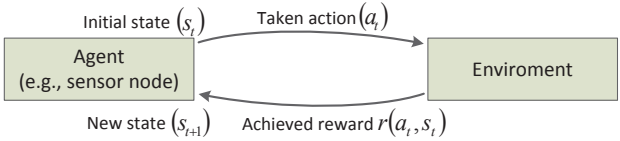}
\par\end{centering}

\caption{\label{fig:q_learning}A Visualization of the Q-learning method.}
\end{figure}

Reinforcement learning enables an agent (e.g., a sensor node) to learn by interacting with its environment. The agent will learn to take the best actions that maximize its long-term rewards by using its own experience. The most well-known reinforcement learning technique is Q-learning \cite{15852369}. As shown in Fig. \ref{fig:q_learning}, an agent regularly updates its achieved rewards based on the taken action at a given state. The future total reward (i.e., the Q-value) of performing an action $a_{t}$ at a given state $s_{t}$ is computed using Eq. (\ref{eq:rl_update}). 

\begin{equation}
Q\left(s_{t+1},a_{t+1}\right)=Q\left(s_{t},a_{t}\right) +\gamma \left(r\left(s_{t},a_{t}\right)-Q\left(s_{t},a_{t}\right)\right)\label{eq:rl_update}
\end{equation}
where $r(s_{t},a_{t})$ denotes the immediate reward of performing an action $a_{t}$ at a given state $s_{t}$, and $\gamma$ is the learning rate that determines how fast learning occurs (usually set to value between 0 and 1). This algorithm can be easily implemented in a distributed architecture like WSNs, where each node seeks to choose actions that are expected to maximize its long term rewards. It is important to note that Q-learning has been extensively and efficiently used in WSN routing problem (e.g., \cite{1181362,4411037,4496872,4496810}). 

\section{\label{sec:functional_challenges}FUNCTIONAL CHALLENGES}

In the design of WSNs, it is important to consider power and memory constraints of sensor nodes, topology changes, communication link failures, and decentralized management. Machine learning paradigms have been successfully adopted to address various functional challenges of wireless sensor networks such as energy aware and real-time routing, query processing and event detection, localization, node clustering and data aggregation.

\subsection{Routing in WSNs}

Designing a routing protocol for WSNs has to consider various design challenges such as energy consumption, fault tolerance, scalability, and data coverage \cite{1368893}. Sensor nodes are provided with limited processing capabilities, small memory and low bandwidth. Traditionally, it is common to formulate a routing problem in wireless sensor networks as a graph $G=(V,E)$, where $V$ represents the set of all nodes, and $E$ represents the set of bidirectional communication channels connecting the nodes. Using this model, the routing problem can be defined as the process of finding the minimum cost path starting at the source vertex, and reaching all destination vertices, by using the available graph edges. This path is actually a spanning tree $T=(V,E)$ whose vertices include the source (i.e., a root node) and destinations (i.e., leaf nodes that do not have any child nodes). Solving such a tree with optimal data aggregation is found to be NP-hard, even when the full topology is known \cite{1030829}.

\begin{figure}[t]
\begin{centering}
\subfloat[\label{fig:routing_1}Original graph.]{\includegraphics[trim=0.5cm 0.5cm 0.5cm 1cm,width=0.3\columnwidth]{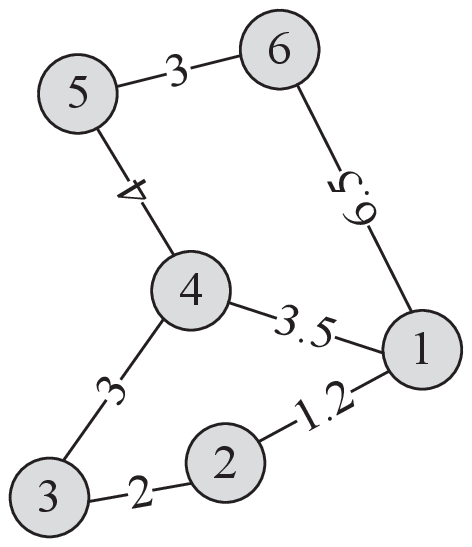}

}\subfloat[\label{fig:routing_2}Traditional routing.]{\includegraphics[trim=0.5cm 0.5cm 0.5cm 1cm,width=0.34\columnwidth]{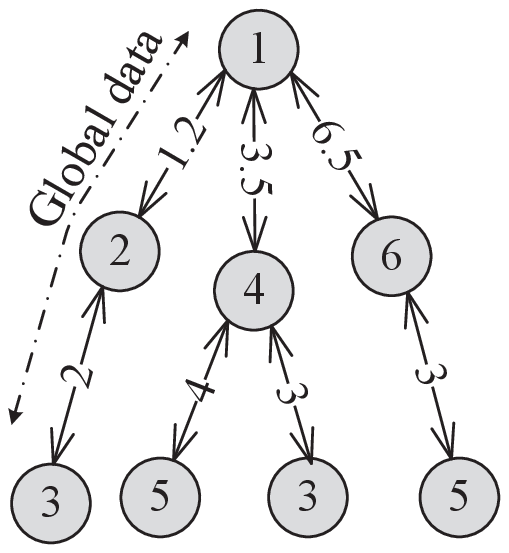}

}\subfloat[\label{fig:routing_3}Simplified problems using machine learning.]{\includegraphics[trim=0.5cm 0.5cm 0.5cm 1cm,width=0.34\columnwidth]{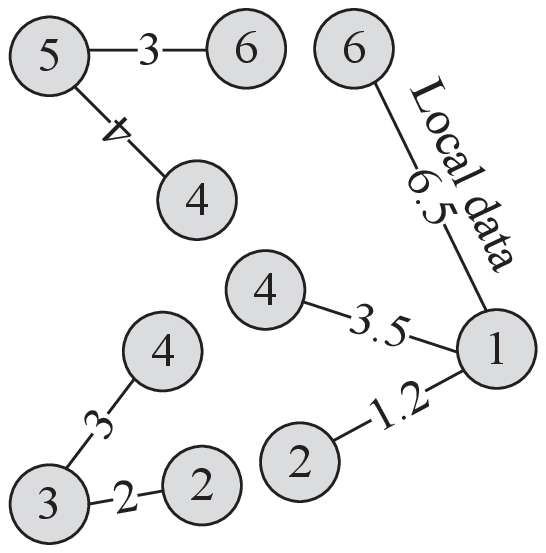}

}
\par\end{centering}

\caption{\label{fig:representation_routing}An example of a sensor network routing problem using a graph along with each path routing cost, traditional spanning tree routing, and the generated sub-problems using machine learning that require only local communication to achieve optimal routing (i.e., require only single-hop neighborhood information exchange).}
\end{figure}

Machine learning allows a sensor network to learn from previous experiences, make optimal routing actions and adapt to the dynamic environment.
The benefits can be summarized as follows:
\begin{itemize}
\item Able to learn the optimal routing paths that will result in energy saving and prolonging the lifetime of dynamically changing WSNs.
\item Reduce the complexity of a typical routing problem by dividing it into simpler sub-routing problems. In each sub-problem, nodes formulate the graph structures by considering only their local neighbors, thus achieving low cost, efficient and real-time routing.
\item Meet QoS requirements in routing problem using relatively simple computational methods and classifiers.
\end{itemize}
Figures \ref{fig:routing_1} and \ref{fig:routing_2} illustrate a simple sensor network routing problem using a graph, and the traditional spanning tree routing algorithm, respectively. To find the optimal routing paths, the network nodes have to exchange their routing information with each other.  In the other side, Figure \ref{fig:routing_3} demonstrates how machine learning reduces the complexity of a typical routing problem by only considering neighboring nodes' information that will be used to predict the full path quality. Each node will independently perform the routing procedures to decide which channels to assign, and the optimal transmission power. As we will discuss in this subsection, such mechanism is proven to provide a near optimal routing decision with a very low computational complexity.

In this subsection, a wide range of machine learning-based routing protocols developed for WSNs are described. Table \ref{tab:Summary-ml-routing} provides
a summary and comparison of these routing protocols. The column \textquotedblleft{}Scalability\textquotedblright{} implies the solutions' capability to route data in large scale networks.

\begin{table*}[t]
\begin{centering}
\caption{\label{tab:Summary-ml-routing}Summary of wireless sensor network routing protocols that adopt machine learning paradigms.}
\begin{tabular}{|>{\centering}m{0.14\textwidth}|>{\centering}m{0.12\textwidth}|>{\centering}m{0.12\textwidth}|>{\centering}m{0.1\textwidth}|c|>{\centering}m{0.09\textwidth}|>{\centering}m{0.1\textwidth}|c|}
\hline 
\textbf{\noun{Routing protocols}} & \textbf{\noun{Topology}} & \textbf{\noun{Machine learning algorithm(s)}} & \textbf{\noun{Overhead}} & \textbf{\noun{Scalability}} & \textbf{\noun{Delay}} & \textbf{\noun{Distributed / centralized}} & \textbf{\noun{QoS }}\tabularnewline
\hline 
\hline 
Distributed regression \cite{1307317} & Flat / multi-hop & kernel linear regression & Low  & Limited & High & Distributed & No\tabularnewline
\hline 
SIR \cite{9120642} & Flat / multi-hop & SOM & High & Limited & Moderate & Hybrid & Yes\tabularnewline
\hline 
Q-MAP multicast \cite{1181362} & Flat /multi-hop & Q-learning & Low & Moderate & High & Distributed & No\tabularnewline
\hline 
RLGR \cite{4411037} & Hierarchical / geographic routing & Q-learning  & Low & Good & Low & Distributed & No\tabularnewline
\hline 
Q-Probabilistic \cite{4496810} & Flat / geographic routing & Q-learning  & Low & Limited & High & Distributed & Yes\tabularnewline
\hline 
FROMS \cite{4496872}  & Flat - multi-hop & Q-learning & High & Limited & Moderate & Distributed & No\tabularnewline
\hline 
\end{tabular}\\
\par\end{centering}
\end{table*}

\subsubsection{Distributed regression framework}

In \cite{1307317}, Guestrin \textit{et al.} introduced a general framework for sensors data modeling. This distributed framework relies on the network nodes for fitting a global function to match their own measurement. The nodes are used to execute a kernel linear regression in the form of weighted components. Kernel functions map the training samples into some feature space to facilitate data manipulation (refer to \cite{Scholkopf:2001:LKS:559923,1315937} for an introduction to kernel methods). The proposed framework exploits the fact that the readings of multiple sensors are highly correlated. This will minimize the communication overhead for detecting the structure of the sensor data. Collectively, these results serve as an important step in developing a distributed learning framework for wireless networks using linear regression methods. The main advantages of utilizing this algorithm are the good fitting results, and the small overhead of the learning phase. However, it cannot learn non-linear and complex functions.

\subsubsection{Data routing using self-organizing map (SOM)}

\begin{figure}[t]
\begin{centering}
\includegraphics[trim=1cm 1cm 1cm 1cm, width=0.75\columnwidth]{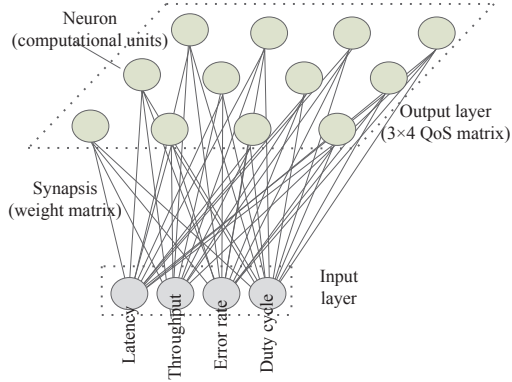}
\par\end{centering}
\caption{\label{fig:sir_som}The SOM construction of the SIR algorithm, where routing link is selected based on the multi-hop path QoS metrics (latency, throughput, error rate, and duty cycle) and the Dijkstra's algorithm \cite{9120642}.}
\end{figure}

Barbancho \textit{et al.} \cite{9120642} introduced ``Sensor Intelligence Routing'' (SIR) by using SOM unsupervised learning to detect optimal routing paths as illustrated in Fig. \ref{fig:sir_som}. SIR introduces a slight modification on the Dijkstra's algorithm to form the network backbone and shortest paths from a base station to every node in the network. During route learning, the second layer neurons compete with each other to reserve high weights in the learning chain. Accordingly, the weights of the winning neuron and its neighboring neurons are updated to further match the input patterns. Clearly, the learning phase is a highly computational process due to the neural network generation task. As a result, it should be performed within a resourceful central station. However, the execution phase does not incur computational cost, and can be run on the network nodes. As a result, this hybrid technique (i.e., a combination of the Dijkstra's algorithm and the SOM model) takes into account the QoS requirements (latency, throughput, packet error rate, and duty cycle) during the process of updating neurons' weights. The main obstacles of applying such an algorithm are the complexity of the algorithm and the overhead of the learning phase in the case that the network's topology and setting change.

\subsubsection{Routing enhancement using reinforcement learning (RL)}

In multicast routing, a node sends the same message to several receivers. Sun \textit{et al.} \cite{1181362} demonstrated the use of Q-learning algorithm to enhance multicast routing in wireless ad hoc networks. Basically, the Q-MAP multicast routing algorithm is designed to guarantee reliable resource allocation. A mobile ad hoc network may consist of heterogeneous nodes, where different nodes have different capabilities. In addition, it is not feasible to maintain a global, up-to-date knowledge about the whole network structure. The multicast routes are determined in two phases. The first phase is ``Join Query Forward'' that discovers an optimal route, as well as updates the Q-values (a prediction of future rewards) of the Q-learning algorithm. The second phase, called ``Join Reply Backward'', creates
the optimal path to allow multicast transmissions. Using Q-learning for multicast routing in mobile ad hoc networks can reduce the overhead for route searching. However, energy efficiency is the key requirement for WSNs, so Q-MAP needs to be modified for WSNs (e.g., considering hierarchical and geographic routing).

The Federal Communications Commission (FCC) has dedicated the frequency band from 3.1 to 10.6 GHz (7,500 MHz of spectrum) for the use of unlicensed ultra-wideband (UWB) communication \cite{1201597}. UWB is a technique for transmitting bulky data for short distances using a wide spectrum of frequency bands with relatively low power. In \cite{4411037}, Dong \textit{et al.} used a similar idea as \cite{1181362} to enhance geographic routing in UWB equipped sensor networks. ``Reinforcement Learning based Geographic Routing'' (RLGR) protocol considers the sensor node energy and delay as metrics for formulating the learning reward function. This hierarchical geographic routing uses the UWB technology for detecting the nodes' locations, where only the cluster heads are equipped with UWB devices. Moreover, each node uses a simple look-up table to maintain the information about its neighbors (as location and energy of the neighbors are needed during network learning). These information are exchanged between nodes using short ``hello'' messages to learn the best routing actions. The main benefit of using reinforcement learning in routing is that it does not require information about the global network structure to achieve an acceptable routing solution.

In \cite{4496810}, Arroyo-Valles \textit{et al.} introduced ``Q-Probabilistic Routing'' (Q-PR), an enhanced geographic routing algorithm for WSNs that learns from previous routing decisions (e.g., to select the routing path that has the highest delivery rate over the past period of time). This protocol differs from RLGR \cite{4411037} in the QoS support. Depending on the importance of messages, expected delivery rate, and the power constraints, Q-PR determines the optimal routes using reinforcement
learning and a Bayesian decision model. This algorithm discovers the next hop during the message routing time (i.e., an on-line opeartion). A Bayesian method is used to handle the decision of transmitting the packets to the set of candidate neighbor nodes, taking into account the data importance, nodes' profiles, and expected transmission and reception energy.

F{\"o}rster and Murphy \cite{4496872} also introduced an enhancement to routing in WSN using reinforcement learning. A novel technique for exchanging node local information as a feedback response to other nodes, named ``Feedback Routing for Optimizing Multiple Sinks in WSN'' (FROMS) is introduced. The main advantage of FROMS is to allow efficient routing from multiple sources to multiple sinks. The Q-values are initialized based on the hop counts to every node in the network. The hop counts can be collected using short ``hello messages'', exchanged between the nodes at earlier stages of the network deployment.
FROMS extends the basic mechanism of RLGR \cite{4411037} by assuming that all nodes can directly communicate with their neighbors.

The key disadvantage of reinforcement learning-based routing algorithms is the limited recognition of future knowledge (i.e., inability to look ahead). Therefore, the algorithms are not suitable for highly dynamic environments as they require a long time to learn optimal routes.

\subsection{Clustering and Data Aggregation}

In large scale energy-constrained sensor networks, it is inefficient to transmit all data directly to the sink \cite{4062839}. One efficient solution is to pass the data to a local aggregator (known as a cluster head) which aggregates data from all the sensors within its cluster and transmits to the sink. This will typically result in energy savings. There are several works that have discussed the optimal selection of the cluster head (i.e., cluster head election process), such as in \cite{1630358,4493846,1420160}. Taxonomy and comparison of classical clustering algorithms are presented in \cite{abbasi2007survey}.

\begin{figure}[t]
\begin{centering}
\includegraphics[trim=1.5cm 1.6cm 1.5cm 0.4cm, width=0.75\columnwidth]{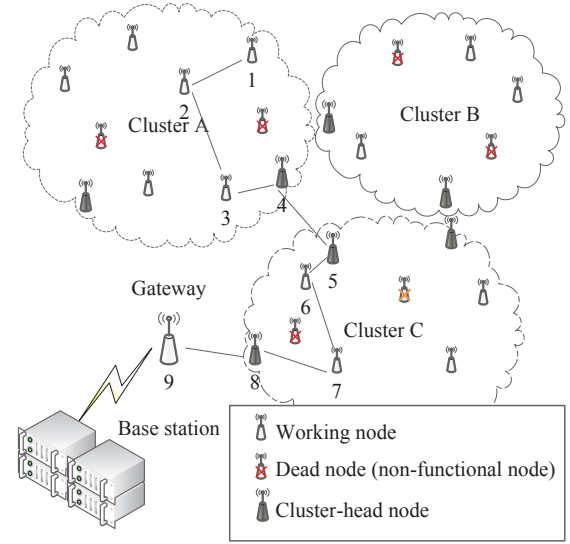}
\par\end{centering}
\caption{\label{fig:data_aggregation_nodes_clustering}Data aggregation example in a clustered architecture, where the nodes are marked as working, dead and cluster heads. }
\end{figure}

Figure \ref{fig:data_aggregation_nodes_clustering} represents the cluster-based data aggregation from sources to a base station in WSNs. In this case, there could be some faulty nodes which must be removed
from the network. Such faulty nodes may generate incorrect readings that could negatively affect the accuracy of the overall operation of the network. Principally, ML techniques improve the operation of node clustering and data aggregation as follows:
\begin{itemize}
\item Usage of machine learning to compress data locally at cluster heads by efficiently extracting similarity and dissimilarity (e.g., from faulty nodes) in different sensors' readings.
\item Machine learning algorithms are employed to efficiently elect the cluster head, where appropriate cluster head selection will significantly reduce energy consumption and enhance the network's lifetime.
\end{itemize}

Table \ref{tab:compassion_ clustering_aggregation} compares data
aggregation and node clustering solutions. The column \textquotedblleft{}Balancing
energy consumption\textquotedblright{} indicates whether the protocol
distributes computationally intensive tasks into all nodes
while considering the remaining energy information. The column ``Topology
aware'' indicates the requirement for full network topology knowledge.

\begin{table*}[t]
\begin{centering}
\caption{\label{tab:compassion_ clustering_aggregation}Compassion of different machine learning-based data aggregation and node clustering mechanisms.}
\begin{tabular}{|>{\centering}m{0.28\textwidth}|>{\centering}m{0.12\textwidth}|>{\centering}m{0.09\textwidth}|>{\centering}m{0.1\textwidth}|>{\centering}p{0.08\textwidth}|c|>{\centering}m{0.08\textwidth}|}
\hline 
\textbf{\noun{Mechanisms}} & \textbf{\noun{Machine learning algorithm(s)}} & \textbf{\noun{Complexity}} & \textbf{\noun{Balancing energy consumption}} & \textbf{\noun{Delay}} & \textbf{\noun{Overhead}} & \textbf{\noun{Topology aware}}\tabularnewline
\hline 
\hline 
Large scale network clustering \cite{4840489} & NNs & Moderate & Yes & High & Low & Yes\tabularnewline
\hline 
Cluster head election \cite{4761982} & DT & Low & Yes & Low & Low & Yes\tabularnewline
\hline 
Gaussian process models for censored sensor readings \cite{4301342}  & \multirow{2}{0.1\textwidth}{\quad{}\quad{}GPR} & Moderate & No & Moderate & Moderate & No\tabularnewline
\cline{1-1} \cline{3-7} 
Adaptive sampling \cite{Kho:2009:DCA:1525856.1525857} &  & High & Yes & High & High & No\tabularnewline
\hline 
Clustering using SOM and sink distance \cite{5489603} & SOM & Moderate & No & High & Moderate & Yes\tabularnewline
\hline 
Online data compression \cite{4024710} & LVQ & High & No & High & High & Yes\tabularnewline
\hline 
Data acquisition using compressive sensing \cite{5425458,5345599} & \multirow{3}{0.1\textwidth}{\quad{}\quad{}PCA} & High & Yes & High & High & Yes\tabularnewline
\cline{1-1} \cline{3-7} 
Transmission reduction \cite{5706781} &  & Moderate & No & High & High & Yes\tabularnewline
\cline{1-1} \cline{3-7} 
Consensus-based distributed PCA \cite{5671089} &  & Moderate & Yes & High & High & No\tabularnewline
\cline{1-1} \cline{3-7} 
Lossy data compression \cite{fenxiong2013data} &  & Moderate & No & Moderate & High & Yes\tabularnewline
\hline 
Collaborative signal processing \cite{985674} & \multirow{2}{0.1\textwidth}{\quad{}\enskip{}k-means} & Low & Yes & Moderate & Moderate & No\tabularnewline
\cline{1-1} \cline{3-7} 
Advanced surveillance systems \cite{4249813} &  & Moderate & Yes & Low & Low & Yes\tabularnewline
\hline 
Role-free clustering \cite{5158454} & Q-learning & Low & No & Low & Low & No\tabularnewline
\hline 
Decentralized learning for data latency \cite{5486214} & RL & Moderate & Yes & Low & Low & No\tabularnewline
\hline 
\end{tabular}
\par\end{centering}
\end{table*}

\subsubsection{Large scale network clustering using neural network}

Hongmei \textit{et al.} \cite{4840489} discussed the development of self-managed clusters using neural networks. This scheme targets the clustering problem in large scale network with short transmissions radii in which centralized algorithms may not work efficiently. However, for large transmission radii, the performance of this algorithm is close to that of centralized algorithms in terms of efficiency and
quality of service.

\subsubsection{Electing a cluster head using decision trees}

Ahmed \textit{et al.} \cite{4761982} applied a decision tree algorithm to solve the cluster head election problem. This approach uses several critical features while iterating the input vector through the decision tree such as distance to the cluster centroids, battery level, the degree of mobility, and the vulnerability indications. The simulation reveals that this scheme enhances the overall performance of cluster head selections when compared to the ``Low Energy Adaptive Clustering Hierarchy'' (LEACH)~\cite{heinzelman2000application} algorithm.

\subsubsection{Gaussian process models for sensor readings}

Gaussian process (GP) is a combination of random variables (stochastic variables) that is parameterized using mean and covariance functions. Ertin \cite{4301342} presented a scheme for initializing probabilistic models of the readings based on Gaussian process regression. Comparatively, Kho \textit{et al.} \cite{Kho:2009:DCA:1525856.1525857} also extended Gaussian process regression to adaptively sample sensor data depending on its importance. Focusing on energy consumption, \cite{Kho:2009:DCA:1525856.1525857} studied a trade-off between computational cost and solution optimality. Broadly speaking, Gaussian process models are preferable in the problems with small training data sets (less than a few thousand samples) and for predicting smooth functions \cite{Rasmussen06gaussianprocesses}. However, WSN designers must consider the high computational complexity of such methods when dealing with large scale networks.

\subsubsection{Data aggregation using self-organizing map (SOM)}

The SOM algorithm is an unsupervised, competitive learning method for mapping from high dimensional spaces to low dimensions. Lee \textit{et al.} \cite{5489603} proposed a novel network architecture called ``Cluster-based self-Organizing Data Aggregation'' (CODA). In this architecture, the nodes are able to classify the aggregated data using a self-organizing algorithm. The winning neuron $j^{*}$, that has a weight vector $w(t)$  closest to the input vector $x(t)$, is defined as:

\begin{equation}
j^{*}=\arg\min_{j}\left\Vert x_{j}(t)-w_{j}(t)\right\Vert ,j=1,...,N\label{eq:wining_node}
\end{equation}
where $N$ represents the number of neurons in the second layer. Further, the winning node and its neighbors are updated as follows:

\begin{equation}
w_{j}\left(t+1\right)=w_{j}\left(t\right)+h\left(t\right)\left(x_{j}\left(t\right)-w_{j}\left(t\right)\right)\label{eq:som_update}
\end{equation}
where $w\left(t\right)$ and $w\left(t+1\right)$ represent the values of a neuron at time $t$ and $t+1$, respectively. In addition, $h\left(t\right)$ is the Gaussian neighborhood function given as:

\begin{equation}
h\left(t\right)=\frac{1}{\sqrt{2\pi}\sigma}\exp\left(-\frac{\left\Vert j^{*}-j\right\Vert ^{2}}{2\sigma^{2}\left(t\right)}\right).\label{eq:neighbor_update}
\end{equation}

Using CODA for data aggregation will result in enhancing the quality of data, saving network energy, and reducing the network traffic.

\subsubsection{Applying learning vector quantization for online data compression}

While the above methods require a complete knowledge about the network topology, some algorithms may not have such a restriction. For example, Lin \textit{et al.} \cite{4024710} introduced a technique called
``Adaptive Learning Vector Quantization'' (ALVQ) to accurately retrieve compressed versions of readings from the sensor nodes. Using data correlation and historical patterns, ALVQ uses the LVQ learning algorithm to predict the code-book using past training samples. The ALVQ algorithm minimizes the required bandwidth during transmission, and enhances the accuracy of original reading recovery from the compressed data. 

The crucial disadvantage of using LVQ for online data aggregation is that dead neurons, that are far away from the training samples, will never take part in the competition. Therefore, it is important to develop algorithms that are robust against outliers. By the same token, LVQ is suitable for representing big data set by few vectors \cite{Kohonen:2001}.

\subsubsection{Data aggregation using principal component analysis}

We begin by introducing two important algorithms that are efficiently used in combination with principal component analysis (PCA) to enhance data aggregation in WSNs. 
\begin{itemize}
\item \textbf{Compressive sensing (CS)} has been recently explored to replace the traditional scheme of ``sample then compress'' with ``sample while compressing''. CS explores sparsity property of signals to
recover the original signal from few random measurements. A simple introduction to CS is provided in \cite{5954192}. 
\item \textbf{Expectation-maximization (EM)} \cite{dempster1977maximum} is an iterative algorithm composed of two steps, i.e., an expectation (E) step and a maximization (M) step. During its E-step, EM formulates the cost function while fixing the current expectation of the system parameters. Subsequently, the M-step recomputes parameters that minimize the estimation error of the cost function.
\end{itemize}
Masiero \textit{et al.} \cite{5425458,5345599} developed a method for estimating distributed observations using few collected samples from a WSN. This solution is based on the PCA technique to produce
orthogonal components used by compressive sensing to reconstruct the original readings. Moreover, this method is independent of the routing protocol due to its ability to estimate data spatial and temporal correlations. Similarly, Rooshenas \textit{et al.} \cite{5706781} applied PCA to optimize the direct transmission of readings to a base station. PCA results in considerable traffic reduction by combining
nodes' collected data into fewer packets. This distributed technique is executed in intermediate nodes to combine all the incoming packets instead of forwarding them to destinations.

Equally important, Macua \textit{et al.} \cite{5671089} introduced distributed consensus-based methods for data compression using PCA and maximum likelihood of the observed data. These methods are ``Consensus-based Distributed PCA'' (CB-DPCA) which relies on exploring the eigenvectors of local covariance matrices, and ``Consensus-based EM Distributed PCA'' (CB-EM-DPCA). The latter uses a distributed EM algorithm. These methods adopt the consensus algorithm \cite{degroot1974reaching} to predict the probability distribution of the data, and hence calculate the global dominant eigenvectors using only local communication parameters (i.e., single hop communications). CB-DPCA and CB-EM-DPCA can be tuned to provide a trade-off between the achieved approximation quality and the communication cost by adjusting the consensus round parameter. For example, to increase the algorithm accuracy, the number of consensus rounds should be increased which will increase the computational  requirements of the algorithm.

Recently, Fenxiong \textit{et al.} \cite{fenxiong2013data} have tackled the problem of data compression using PCA by transforming the data from a high dimensional space to a lower one. The data is collected over time, and then it is transmitted from each node to its corresponding cluster head. At the cluster head, the data matrix is compressed to eliminate the data redundancy.  The data compression is achieved by ignoring principal components through which the data has the least variation values.

High computational requirement is the main issue of PCA-based data aggregation solutions. Other than increasing throughput, these solutions elegantly cope with the high dimensionality of collected data by keeping only important information (data dimensionality reduction).

\subsubsection{Collaborative data processing through k-means algorithm}

Li \textit{et al.} \cite{985674} addressed the fundamental concepts for distributed detection and tracking of a single target using sensor networks. ``Collaborative Signal Processing'' (CSP) is a framework for information gathering from the monitored environment. Additionally, this algorithm can track multiple targets using classification techniques such as SVM and k-nearest neighbors.

Classical surveillance systems have to collect massive data from surveillance cameras. Together with the requirement of highly complex computation and analysis process, this introduces the need for more practical methods. Therefore, Tseng \textit{et al.} \cite{4249813} proposed ``Integrated Mobile Surveillance and Wireless Sensor System'' (iMouse) which adopts powerful mobile sensors
to enhance traditional surveillance systems. iMouse divides the monitored sites into a number of clusters using the k-means unsupervised learning algorithm. Each cluster will be repeatedly monitored by only one mobile
sensor.

Although these ideas (using k-means for data processing) are appealing because of the straightforward implementations and low complexity, they are still sensitive to outliers and to initial seed selections.

\subsubsection{Role-free clustering}

In \cite{5158454}, F{\"o}rster and Murphy introduced the WSN cluster formulation method called ``Role-Free Clustering with Q-Learning for Wireless Sensor Networks'' (CLIQUE). Instead
of performing an election process, CLIQUE enables each node to investigate its ability to function as a cluster-head node. This is achieved through the use of Q-learning algorithm in combination with some dynamic network parameters such as energy levels.

\subsubsection{Decentralized learning for data latency}

Mihaylov \textit{et al.} \cite{5486214} addressed the problem of high data latency in random topology sensor networks using reinforcement learning. Each node executes the learning algorithm locally to optimize
the data aggregation without the need for a central control station. Consequently, the efficiency of the whole network is enhanced with smaller learning transmission overhead. The approach saves the node energy budget during data gathering process, and hence prolong the network lifetime.

\subsection{Event Detection and Query Processing}

\begin{figure}[t]
\begin{centering}
\includegraphics[trim=0.4cm 1.2cm 0.4cm 0cm, width=1\columnwidth]{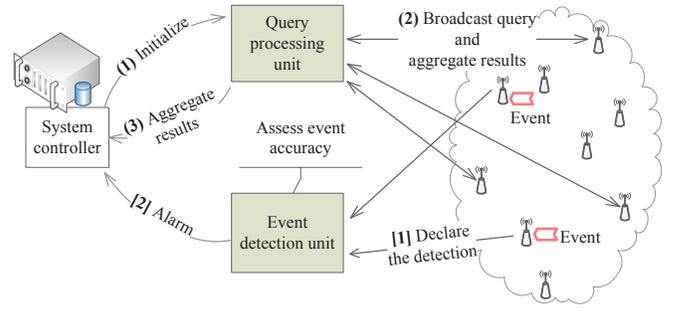}
\par\end{centering}

\caption{\label{fig:event_detection_query_processing}Event detection and query processing enhancement using machine learning methods by assessing event validity and delimiting queried areas. System controller initiates query that is spread by the query processing unit to intended nodes. In contrast, events are detected by nodes to monitor specific signs within the monitored area.
}
\end{figure}

Event detection and query processing are considered to be functional requirements of any large scale sensor network. This introduces the
need for trustworthy event scheduling and detection with minimal human intervention. Monitoring in WSNs can be classified as: event-driven, continuous, or query-driven \cite{1368893}. Figure \ref{fig:event_detection_query_processing} illustrates event detection and query processing operations in WSNs. Fundamentally, machine learning offers solutions to restrict query areas and assess event validity for efficient event detection and query processing mechanisms. This adoption will result in the following benefits:
\begin{itemize}
\item Learning algorithms enable the development of efficient event detection mechanisms with limited requirements of storage and computing resources. Besides they are able to assess the accuracy of such events using simple classifiers.
\item Machine learning facilitates the development of effective query processing techniques for WSNs, that determine the search regions whenever a query is received without flooding the whole network.
\end{itemize}

The design of effective event detection and query processing solutions has recently received increased attention from WSNs research community. The simplest techniques rely on defining a strict threshold value for the sensed phenomenon and alarming the system manager of any violations. However, in most recent applications of WSNs, event and query processing units are often complicated and require more than a predefined threshold value. One such emerging technique is to use machine learning to develop advanced event detection and query processing solutions. Table \ref{tab:comparison_query_processing} presents a comparison of functional aspects of different machine learning-based event detection and query processing solutions for WSNs.

\begin{table*}[t]
\begin{centering}
\caption{\label{tab:comparison_query_processing}Comparison of functional aspects of different machine learning-based event detection and query processing solutions for WSNs.}

\begin{tabular}{|>{\centering}m{0.26\textwidth}|>{\centering}m{0.12\textwidth}|>{\centering}m{0.12\textwidth}|>{\centering}m{0.12\textwidth}|>{\centering}m{0.26\textwidth}|}
\hline 
\textbf{\noun{Approaches}} & \textbf{\noun{Machine learning algorithm(s)}} & \textbf{\noun{Data Delivery Models}} & \textbf{\noun{Complexity}} & \textbf{\noun{Characteristics}}\tabularnewline
\hline 
\hline 
Event region detection \cite{1261832}  & \multirow{2}{*}{\enskip{}Bayesian} & \multirow{2}{0.1\textwidth}{\enskip{}Event-driven} & Low & Fault-tolerant event region detection\tabularnewline
\cline{1-1} \cline{4-5} 
Activity recognition \cite{5932548} &  &  & Moderate & Real-time activity recognition\tabularnewline
\hline 
In-network query processing \cite{1541008} & \multirow{2}{*}{k-NN} & \multirow{2}{0.1\textwidth}{\enskip{}Query-driven} & Low & In-network query processing\tabularnewline
\cline{1-1} \cline{4-5} 
Query processing in 3D space \cite{5698542}  &  &  & Moderate & Enhance 3D space query processing\tabularnewline
\hline 
Forest fire detection \cite{1544272} & NNs & Event-driven & Moderate & Real-time and lightweight forest fire detection\tabularnewline
\hline 
Distributed event detection \cite{5702151}  & DT & Event-driven & Low & Disaster prevention system\tabularnewline
\hline 
Query optimization \cite{5486201} & PCA & Query-driven & High & Query optimization and dimensionality reduction\tabularnewline
\hline 
\end{tabular}
\par\end{centering}
\end{table*}

\subsubsection{Event recognition through Bayesian algorithms}

Krishnamachari and Iyengar \cite{1261832} investigated the use of WSNs for detecting environmental phenomenon in a distributed manner. Readings will be considered as faulty if their values exceed a specific
threshold. This study employs decentralized Bayesian learning that detects up to 95 percent of the faults, and will result in recognizing the event region. It is important to note that Chen \textit{et al.} \cite{1471678} provided corrections to several errors related to the distributed Bayesian algorithms that have been derived in \cite{1261832}. In summary, these corrections result in enhanced error and performance calculations for the distributed Bayesian algorithm proposed in \cite{1261832}.

\begin{figure}[t]
\begin{centering}
\includegraphics[trim=0.5cm 0.5cm 0.5cm 0.5cm, width=1.0\columnwidth]{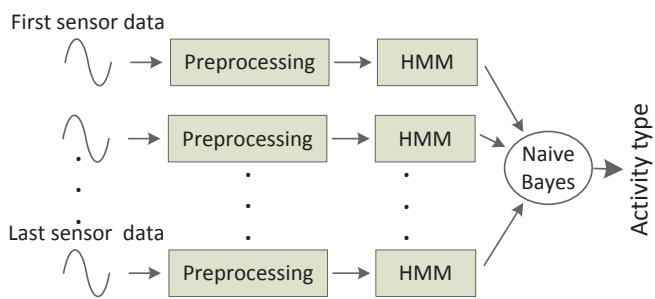}
\par\end{centering}

\caption{\label{fig:hmm_naive_activity}Human activity recognition using the hidden Markov model and the naive Bayes classifier \cite{5932548}.}
\end{figure}

Additionally, Zappi \textit{et al.} \cite{5932548} presented a real-time approach for activity recognition using WSNs that accurately detects body gesture and motion. Initially, the nodes, that are spread throughout the body, detect the organ motion using an accelerometer sensor with three axis measurements (positive, negative and null), where these measurements are used by a hidden Markov model (HMM) to predict the activity at each sensor.  Sensor activation and selection rely on the sensor's potential contributions in classifier accuracy (i.e., select the sensors that provide the most informative description of the gesture). To generate a final gesture decision, a naive Bayes classifier is used to combine the independent node predictions so as to maximize the posterior probability of the Bayes theorem. The architecture of the proposed system is shown in Fig. \ref{fig:hmm_naive_activity}. 

\subsubsection{Forest fire detection through neural network}

WSNs were actively used in fire detection and rescue systems  (see \cite{4017702} and references therein for requirements and challenges of such systems). Moreover, the use of WSNs for forest fire detection can achieve better performance than using satellite-based solutions while costing much less. Yu \textit{et al.} \cite{1544272} presented a real-time forest fire detection scheme based on a neural network method. Data processing will be distributed to cluster heads, and only important information will be aggregated to a final decision maker. Although the idea is creative and beneficial to the environment, the classification task and system core are hardly interpretable when introducing such systems to decision makers.

\subsubsection{Query processing through k-nearest neighbors}

K-nearest neighbor query is considered as a highly effective query processing technique in WSNs. For example, Winter \textit{et al.} \cite{1541008} developed an in-network query processing solution using the k-nearest neighbor algorithm, namely the ``K-NN Boundary Tree'' (KBT) algorithm. Each node that is aware of its location will determine its k-NN search region whenever a query is received from the application manager.

Correspondingly, Jayaraman \textit{et al.} \cite{5698542} extended the query processing design of \cite{1541008}. ``3D-KNN'' is a query processing scheme for WSNs that adopts the k-nearest neighbor algorithm. This approach restricts the query region to bound at least k-nearest nodes deployed within a 3D space. In addition, signal-to-noise ratio (SNR) and distance measurements are used to refine the k-nearest neighbor.

The primary concerns of such k-NN-based algorithms for query processing are the requirement of large memory footprint to store every collected sample and the high processing delay in large scale sensor networks.

\subsubsection{Distributed event detection for disaster management using decision tree}

Bahrepour \textit{et al.} \cite{5702151} developed decision tree-based event detection and recognition for sensor network disaster prevention systems. The main application of this decentralized mechanism is the fire detection in residential areas. Most noteworthy, the final event detection decision is made by using a simple vote from the highest reputation nodes.

\subsubsection{Query optimization using principal component analysis (PCA)}

\begin{figure}[t]
\begin{centering}
\includegraphics[trim=1cm 1cm 1cm 1cm, width=0.85\columnwidth]{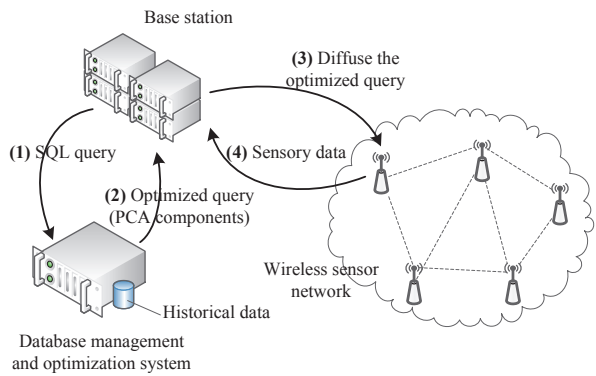}
\par\end{centering}

\caption{\label{fig:pca_query}Workflow of the query optimization and reduction system using PCA proposed in \cite{5486201}}
\end{figure}

Malik \textit{et al.} \cite{5486201} optimized traditional query processing in WSNs using data attributes and PCA, thus reducing the overhead of such a process. PCA has been used to dynamically detect important attributes (i.e., dominant principal components) among the whole correlated data set. Figure \ref{fig:pca_query} shows the workflow of the proposed algorithm in four fundamental steps. In Step 1, the structured query language (SQL) request, which contains the human intelligible attributes, is sent to the database management and optimization system. At the database management and optimization system, the original query is optimized where the high-variance components are extracted from historical data using the PCA algorithm (Step 2). Then, the optimized query is diffused to the wireless sensor network to extract the sensory data as shown in Steps 3 and 4, respectively. Later, the original attributes (i.e., human intelligible attributes) can be extracted from the optimized attributes by reversing the process of PCA. 

As a result, this algorithm guarantees 25 percent improvement in energy saving of the network nodes while achieving 93 percent of accuracy rates. However, this enhancement is at the cost of accuracy of the collected data (as some of the data components will be ignored). Therefore, this solution may not be ideal for the applications with high accuracy and precision requirements.

\subsection{Localization and Objects Targeting}

Localization is the process of determining the geographic coordinates of network's nodes and components. Position awareness of sensor nodes is an important capability, since most sensor network operations are typically based on the location \cite{4343996}. In most large scale systems, it is financially infeasible to use global positioning system (GPS) hardware in each node for this purpose. Moreover, GPS service may not be available in the observed environment (e.g., indoor). Relative location measurement is sufficient for certain uses. However, by using the absolute locations for a small group of nodes, relative locations can be transformed into absolute ones \cite{56532056}. In order to enhance the performance of proximity based localization, additional measurements relying on distance, angle or a hybrid of them can be used. Distance measurements can be obtained by utilizing various techniques such as RSSI, TOA, and TDOA. Furthermore, angle of the received signal can be measured using compasses
or special smart antennas \cite{Nasipuri:2002:DBL:570738.570754}. A valuable introduction about the basics of different range-based localization techniques is provided in \cite{8547966}.

\begin{figure}[t]
\begin{centering}
\includegraphics[trim=1cm 0.5cm 1cm 0.5cm, width=0.65\columnwidth]{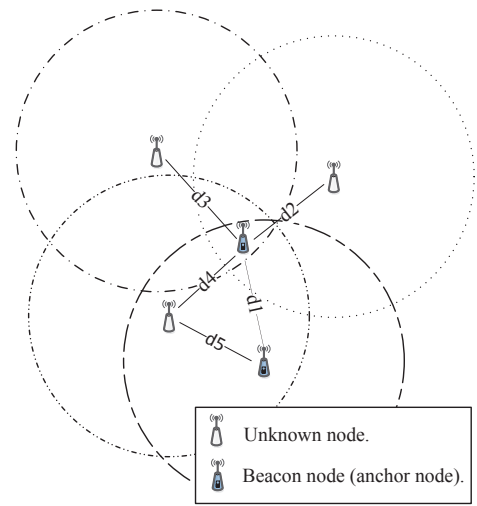}
\par\end{centering}

\caption{\label{fig:localization_problem}Localization using few beacon nodes by utilizing machine learning algorithms and other signal strength indicators (reformulated from \cite{4343996}).}
\end{figure}

Sensor nodes may encounter changes in their location after deployment (e.g., due to movement). The benefits of using machine learning algorithms in sensor node localization process can be summarized as follows:
\begin{itemize}
\item Converting the relative locations of nodes to absolute ones using few anchor points. This will eliminate the need for range measurement hardware to obtain distance estimations.
\item In surveillance and object targeting systems, machine learning can be used to divide the monitored sites into a number of clusters, where each cluster represents specific location indicator.
\end{itemize}
We begin by defining some terms that are widely used in WSN localization literature, as illustrated in Figure \ref{fig:localization_problem}.
\begin{itemize}
\item \textbf{Unknown node} is a node that cannot determine its current location.
\item \textbf{Beacon node (or anchor node)} is any node that is able to recognize its location by using positioning hardware or from its manual placement. In most systems, the beacon node is used as a reference point to estimate the coordinates of other unknown nodes.
\item \textbf{Received signal strength indication (RSSI)} is an indicator of the received signal strength, used to represent transmission performance or distance.
\end{itemize}

Next, we discuss some seminal WSN localization techniques that use machine learning and summarize our reviews in Table \ref{tab:summary-localization}.

\begin{table*}[t]
\begin{centering}
\caption{\label{tab:summary-localization}Summary of localization algorithms in WSNs that adopt machine learning concepts and their prime advantages. The column ``Applications'' specifies the targeted application(s) of the proposed solution (either general-purpose or a specific application).}

\begin{tabular}{|>{\centering}m{0.24\textwidth}|>{\centering}m{0.12\textwidth}|>{\centering}m{0.12\textwidth}|>{\centering}m{0.12\textwidth}|>{\centering}m{0.10\textwidth}|>{\centering}m{0.15\textwidth}|}
\hline 
\textbf{\noun{Studies description}} & \textbf{\noun{Machine learning algorithm(s)}} & \textbf{\noun{Complexity}} & \textbf{\noun{Distributed / centralized}} & \textbf{\noun{Beacon / Beacon-less}} & \textbf{\noun{Applications}}\tabularnewline
\hline 
\hline 
Bayesian node localization \cite{4518167} & \multirow{2}{0.1\textwidth}{\quad{}Bayesian} & Moderate & Centralized & Beacon & General-purpose \tabularnewline
\cline{1-1} \cline{3-6}
Location-aware activity recognition \cite{5161349} &  & Moderate & Centralized & Beacon & Smart homes\tabularnewline
\hline 
Localization based on NNs \cite{Shareef:2008:LUN:1361492.1361497}  & \multirow{3}{0.1\textwidth}{\quad{}\quad{}NNs} & High & Centralized & Beacon & General-purpose\tabularnewline
\cline{1-1} \cline{3-6}
Soft localization \cite{Yun:2009:SCA:1508324.1508595} &  & Moderate & Distributed & Beacon & General-purpose\tabularnewline
\cline{1-1} \cline{3-6}
Localization based on NNs \cite{6328975} &  & High & Distributed & Beacon & General-purpose\tabularnewline
\hline
Area localization \cite{5487032} & \multirow{2}{0.1\textwidth}{\quad{}\quad{}SVM} & Moderate & Distributed & Beacon & General-purpose \tabularnewline
\cline{1-1} \cline{3-6}
Localization using SVM \cite{4384476} &  & Moderate & Distributed & Beacon & General-purpose\tabularnewline
\hline 
Localization using SVR \cite{5506589} & SVR & Moderate & Distributed & Beacon & General-purpose\tabularnewline
\hline 
Target classification and information fusion \cite{4912210} & DT & Low & Distributed & Beacon & Acoustic WSNs \tabularnewline
\cline{1-1} \cline{3-6}
Underwater surveillance system \cite{Cayirci2006431} &  & Moderate & Centralized & Beacon & Submarine surveillance\tabularnewline
\hline 
Sensor placements \cite{Krause:2008:NSP:1390681.1390689} & GP & Low & Distributed & Beacon & Node deployment\tabularnewline
\hline 
Spatial Gaussian process regression \cite{6218781} & GPR & Moderate & Distributed & Beacon & Collective node motion \tabularnewline
\hline 
Localization in 2D space \cite{4381576}  & \multirow{3}{0.1\textwidth}{\quad{}\quad{}SOM} & Moderate & Distributed & Beacon & Large scale WSNs \tabularnewline
\cline{1-1} \cline{3-6} 
Localization using SOM \cite{Giorgetti:2007:WLU:1236360.1236399} &  & Low & Centralized & Beacon-less & General-purpose\tabularnewline
\cline{1-1} \cline{3-6}
Distributed localization \cite{4648079} &  & Moderate & Distributed & Beacon-less & General-purpose\tabularnewline
\hline 
Path determination \cite{6185486} & RL & Low & Distributed & Beacon & Mobile nodes\tabularnewline
\hline 
\end{tabular}
\par\end{centering}
\end{table*}

\subsubsection{Bayesian node localization}

Morelande \textit{et al.} \cite{4518167} used a Bayesian algorithm to develop a localization scheme for WSNs using only few anchor points. This study focuses on the enhancement of progressive correction \cite{musso2001improving}, which is a method for predicting samples from likelihoods to get closer to the posterior likelihood. The proposed algorithm is efficiently applicable for node localization in large scale systems (i.e., networks with a few thousands of nodes). The idea of using the Bayesian algorithm for localization is appealing as it can handle incomplete data sets by investigating prior knowledge and probabilities.

\subsubsection{Robust location-aware activity recognition}

Lu and Fu \cite{5161349} addressed the problem of sensor and activity localization in smart homes. The activities of interest include using the phone, listening to the music, using the refrigerator, studying, etc. In such applications, designers need to comply with both human and environment constraints in a convenient and easily operated way. The proposed framework, named ``Ambient Intelligence Compliant Object'' (AICO), facilities the human interaction with the home electric devices in a more intelligent manner (e.g., automatic power supply management). At its core, AICO uses multiple naive Bayes classifiers to determine the resident's current location and evaluate the reliability of the system by detecting any malfunctioned sensors. Although this method provides a robust mechanism for localization, it is still application-dependent and the designers must predefine a set of supported activities in advance. This is because the used learning features are selected and evaluated manually depending on the activities and the domain of interest. To overcome this limitation in this centralized system, we recommend investigating unsupervised machine learning algorithms for automatic feature extraction such as the deep learning methods \cite{bengio2009learning} and the non-negative matrix factorization algorithm \cite{6165290}.

\subsubsection{Localization based on neural network}

Shareef \textit{et al.} \cite{Shareef:2008:LUN:1361492.1361497} compared three localization schemes that are based on different types of neural networks. In particular, this study considers WSN localization using multi-layer perceptron (MLP), radial basis function (RBF), and recurrent neural networks (RNN). In summary, the RBF neural network results in the minimum error at the cost of high resource requirements. In contrast, MLP consumes the minimum computational and memory resources.

Likewise, Yun \textit{et al.} \cite{Yun:2009:SCA:1508324.1508595} adopted a similar design, in which two classes of algorithms for sensor node localization using RSSI from anchor nodes are proposed. The first
class utilizes the fuzzy logic system and genetic algorithm. In the second class, the neural network is adopted to predict the sensor location by using RSSI measurements from all anchor nodes as an input vector. In the same way, Chagas \textit{et al.} \cite{6328975} applied neural networks for WSNs localization with RSSI as an input to the learning network.

The main advantage of these NNs-based localization algorithms is their ability to provide coordinates in the form of continuous-valued vectors (e.g., coordinates in 3D space). However, unlike statistical
or Bayesian alternatives, neural network is a non-probabilistic method. This fact limits the designers' certainty about precision of unknown node's predicted coordinates, and hence restricts their ability to
manage the cost of localization errors.

\subsubsection{Localization using support vector machine (SVM)}

The SVM technique has been widely used for node localization in WSNs, where having a self-positioning device to each sensor is infeasible. As an illustration, Yang \textit{et al.} \cite{5487032} developed a mobile node localization scheme by employing SVM and connectivity information capabilities. In its initial step, the proposed method has to detect node movement using their radio frequency oscillation such as RSSI metric. For movement detection, SVM will be executed to provide the new location.

Similar to \cite{5487032}, Tran and Nguyen \cite{4384476} proposed ``Localization Based on Support Vector Machines'' (LSVM) method for node localization in WSNs. To achieve its design goals and given an appropriate training data, LSVM adopts several decision metrics such as connectivity information and indicators. Even though LSVM offers distributed localization in a fast and effective manner, its performance is still sensitive to outliers in training samples.

\subsubsection{Localization using support vector regression (SVR)}

Limited resources and high data dimensionality impede the wide adoption of SVR learning in WSNs. Therefore, Kim \textit{et al.} \cite{5506589} developed the idea of using lightweight implementation of SVR by dividing the original regression problem into several sub-problems. Basically, the algorithm starts by dividing the network into a set of sub-networks, thus a small number of data has to be processed by each regression algorithm (i.e., SVR's sub-predictors). Then, the learned hypothesis models of the sub-predictors are combined together using a customized ensemble combination technique. Thus, in addition to its low computational requirements and robustness against noisy data, this solution converges to the preferred solution with low computational requirement.

\subsubsection{Decision tree-based localization}

Based on decision tree learning, Merhi \textit{et al.} \cite{4912210} developed an acoustic target localization method for WSNs. Exact locations of targets are determined using the time difference of arrival
(TDOA) metric in a spatial correlation decision tree. Also, this work proposed the design of ``Event Based MAC'' (EB-MAC) protocol, that enables event-based localization and targeting
in acoustic WSNs. The proposed framework was implemented using a MicaZ board that supports ZigBee 802.15.4 specifications for personal area networks. 

Using the GPS functionality to support localization in underwater wireless sensor network's applications may not be feasible due to the propagation limitation of the GPS signal through water \cite{tan2011survey}. Erdal \textit{et al.} \cite{Cayirci2006431} developed a system for submarine detection in underwater surveillance systems, so that a randomly deployed node finds its location in the 3D space based on beacon node coordinates. Each monitoring unit consists of a sensor that is fixed with a cable to a surface buoy. Data is collected using the buoys, where they are transmitted to the central processing unit. At the central unit, a decision tree classifier is used to recognize any submarines in the monitored sites.

\subsubsection{Sensor placements through Gaussian processes}

Krause \textit{et al.} \cite{Krause:2008:NSP:1390681.1390689} provided an optimized solution to sensor placement in applications with spatially correlated data such as temperature monitoring systems. One interesting feature of this solution is the development of a lazy learning scheme based on Gaussian process model for the investigated phenomenon. Lazy learning algorithms store training samples and delay the major processing task until a classification request is received. Moreover, this solution aims to achieve robustness against node failures and model ambiguity when choosing optimal locations for sensors.

\subsubsection{Spatial Gaussian process regression}

Gu and Hu \cite{6218781} developed a distributed protocol for collective node motion. This approach employs distributed Gaussian process regression (DGPR) to predict optimal locations for mobile nodes' movements. Traditional Gaussian process regression (GPR) algorithms have computational complexity of $\mathcal{O}(N^{3})$, where $N$ is the size of samples. However, this solution adopts a sparse Gaussian process regression algorithm to reduce such computational complexity. Each node will execute the regression algorithm independently using only spatiotemporal information from local neighbors.

\subsubsection{Localization using self-organizing map (SOM)}

Given some anchor positions, Paladina \textit{et al.} \cite{4381576} introduced the SOM-based positioning solution for WSNs consisting of thousands of nodes. The proposed scheme is executed in each node with a simple SOM algorithm that consists of a 3x3 input layer connected to the 2 neurons of the output layer. In particular, the input layer is formulated using the spatial coordinates of 8 anchor nodes surrounding the unknown node. After a sufficient training, the output layer is used to represent the unknown node's spatial coordinates in a 2D space. The main disadvantage of this scheme is that the nodes should be distributed uniformly and equally spaced throughout the monitored area.

Unlike traditional methods that require absolute locations of a few nodes to find the positions of the unknown nodes, Giorgetti \textit{et al.} \cite{Giorgetti:2007:WLU:1236360.1236399} introduced a localization algorithm that is only based on connectivity information and the SOM algorithm. The developed method is highly suitable for networks with limited resources, as it does not require a GPS-enabled device. However, since this is a centralized algorithm, each node transmits the information of its neighbors to the central processing unit to determine the adjacency matrix and hence the node's location. Similarly, Hu and Lee \cite{4648079} presented a scheme that provides node localization service in WSNs without the need for anchor nodes. The algorithm is based on SOM, and it operates efficiently for any number of nodes. The contribution of \cite{4648079} over \cite{Giorgetti:2007:WLU:1236360.1236399} is that the proposed algorithm distributes the computation tasks to all nodes in the network, which eliminates the needs for a central unit and minimizes the transmission overhead of the algorithm.

\subsubsection{Path determination using reinforcement learning}

Li \textit{et al.} \cite{6185486} developed a reinforcement learning-based localization method for WSNs, called ``Dynamic Path determination of Mobile Beacons'' (DPMB), suitable for real-time management of the mobile beacons. The mobile beacon (MB), which is aware of the physical location during its movement, will be used to determine the positions of large number of sensor nodes. In brief, the states of the Q-learning algorithm are used to represent the different positions of the MB, and the algorithm target is to cover all the sensors in the monitored area (i.e., all the sensors should hear a location update message from the MB at some stages). The entire operation will be run in the mobile beacon, and hence, this will save the resources of the unknown nodes. However, as a centralized method, the entire system will fail in the event of mobile beacon malfunctions.

\subsection{Medium Access Control (MAC)}

In WSNs, a number of sensors cooperate to efficiently transfer data. Therefore, designing MAC protocols for WSNs poses different challenges from typical wireless networks, as well as energy consumption and latency \cite{6328420}. Also, the duty cycle (i.e., fraction of time that a sensor node is active) of the node has to be controlled to conserve energy. Therefore, the MAC protocols have to be modified to support efficient data transmission and reception of the sensor nodes. A comprehensive survey of MAC protocols in WSNs is provided in \cite{5451759}.

Recently, machine learning methods have been used to enhance the performance of MAC protocols in WSNs. Specifically, this is achieved through the following points:
\begin{itemize}
\item Machine learning can be used to adaptively determine the duty cycle of a node using the transmission history of the network. In particular, the nodes, which are able to predict when the other nodes' transmissions will finish, can sleep in the meantime and wake up (to transmit data) just when the channel is expected to be idle (i.e., when no other node is transmitting). For WSNs, many factors, such as energy consumption and latency, are more important than fairness when designing MAC protocols.
\item Achieving secured data transmission by combining the concepts of machine learning and MAC protocols. Such MAC layer security schemes are independent of the proposed application and are able to iteratively learn sporadic attack patterns.
\end{itemize}

Table \ref{tab:comparison-ml-mac} gives a brief comparison between MAC protocols reviewed in this subsection. The column ``Synchronization'' indicates whether the protocol assumes that time synchronization is
achieved externally, and ``Adaptivity to changes'' indicates the ability to handle topology changes such as nodes failure.

\begin{table*}[t]
\begin{centering}
\caption{\label{tab:comparison-ml-mac}Comparison of MAC protocols.}

\begin{tabular}{|>{\centering}m{0.18\textwidth}|>{\centering}m{0.12\textwidth}|>{\centering}m{0.09\textwidth}|c|c|>{\centering}m{0.1\textwidth}|c|}
\hline 
\textbf{\noun{MAC protocols}} & \textbf{\noun{Machine learning algorithm(s)}} & \textbf{\noun{Complexity}} & \textbf{\noun{Category}} & \textbf{\noun{Type}} & \textbf{\noun{Adaptivity to changes}} & \textbf{\noun{Synchronization}}\tabularnewline
\hline 
\hline 
Bayesian Statistical Modeling \cite{5612365} & Bayesian & Moderate & Contention\textendash{}Based & CSMA & Good & No\tabularnewline
\hline 
Broadcast scheduling \cite{Shen2008900} & \multirow{2}{0.12\textwidth}{\quad{}\quad{}\quad{}NNs} & Low & Contention\textendash{}Free & TDMA & Weak & Yes\tabularnewline
\cline{1-1} \cline{3-7} 
NN-based secure MAC \cite{Kulkarni:2009:NNB:1704555.1704768} &  & Moderate & Contention\textendash{}Based & CSMA/CA & Average & No\tabularnewline
\hline 
RL-MAC \cite{Liu:2006:RRL:1359189.1359190} & \multirow{2}{0.12\textwidth}{\quad{}\quad{}\quad{}\,RL} & Moderate & Contention\textendash{}Based & CSMA & Good & Yes\tabularnewline
\cline{1-1} \cline{3-7} 
ALOHA-QIR \cite{6328420} &  & Low & Contention\textendash{}Free & Slotted Aloha & Weak & Yes\tabularnewline
\hline
Adaptive MAC layer \cite{sha2013self} & DT & High & Hybrid & Hybrid & Good & Yes\tabularnewline
\hline 
\end{tabular}\\
\smallskip{}

\par\end{centering}
\end{table*}

\subsubsection{Bayesian statistical model for MAC}

Kim and Park \cite{5612365} presented a contention-based MAC protocol for managing active and sleep times in WSNs. Instead of continuously sensing the medium, this scheme utilizes a Bayesian statistical model to learn when the channel can be allocated, and hence save network energy. Furthermore, in its basic design, this scheme is targeted for CSMA contention-based protocols such as ``Sensor MAC'' (S-MAC)
\cite{1019408}, and ``Timeout MAC'' (T-MAC) \cite{vanDam:2003:AEM:958491.958512}.

\subsubsection{Neural network-based MAC}

Time division multiple access (TDMA)-based protocols employ periodic time frames to separate the medium access of different nodes. This process requires a central unit to broadcast a transmission schedule in case of topology changes. Shen and Wang \cite{Shen2008900} proposed a solution to broadcast the transmission schedule in TDMA using a fuzzy hopfield neural network (FHNN) technique. Time slots are distributed among the nodes in a network while maximizing the cycle length, preventing any potential transmission collisions and reducing the processing time.

In the same way, Kulkarni and Venayagamoorthy \cite{Kulkarni:2009:NNB:1704555.1704768} presented an innovative CSMA-based MAC solution, that can prevent denial-of-service (DoS) attacks in WSNs. Denial-of-service is a type of attacks that generates huge useless traffic (i.e., flood the network), thus preventing the delivery of useful data. In such cases, attackers exploit the limitations of WSNs such as limited
bandwidth and buffering capabilities. The proposed solution is based on neural network learning to prevent flooding the network with untruthful data traffic by investigating the network properties and variations such as packet request rate and average packet waiting time. Consequently, the MAC layer will be blocked if the neural network output exceeds a predefined threshold level. More importantly, only nodes in the affected sites will be blocked, as this solution is designed to work in a distributed manner.

\subsubsection{Duty cycle management using reinforcement learning}

Liu and Elhanany \cite{Liu:2006:RRL:1359189.1359190} employed a reinforcement learning technique to introduce RL-MAC, an adaptive MAC protocol for WSNs. Basically, RL-MAC reduces energy usage and increases throughput by optimizing the duty cycle of the network node. Similar to S-MAC \cite{1019408} and T-MAC \cite{vanDam:2003:AEM:958491.958512}, RL-MAC synchronizes node's transmission on a common schedule in a frame-based structure. RL-MAC adaptively determines the slot length, duty cycle and transmission active time depending on the traffic load and the channel bandwidth.

\begin{figure}[t]
\begin{centering}
\includegraphics[trim=3cm 3cm 0.5cm 0.5cm, width=0.85\columnwidth]{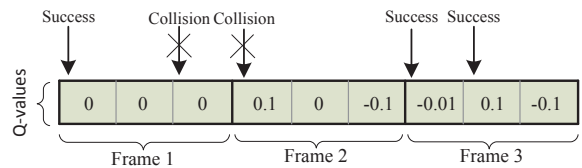}
\par\end{centering}

\caption{\label{fig:aloha_qir}An example of the Q-values of a node over three frames in a WSN that employs ALOHA-QIR to manage medium access \cite{Liu:2006:RRL:1359189.1359190}.}
\end{figure}

Similarly, Chu \textit{et al.} \cite{6328420} integrated slotted ALOHA and Q-Learning algorithms to introduce a new MAC protocol for WSNs, called ``ALOHA and Q-Learning based MAC with Informed
Receiving'' (ALOHA-QIR). ALOHA-QIR inherits the features of both ALOHA and Q-Learning to achieve the benefits of simple design, low resource requirements and low collision probability. During their transmission frames, nodes broadcast their future transmission allocation such that other nodes can sleep during reserved frame. The Q-value map in each node represents the willingness for slot reservation, where the node with higher Q-value will attain the right of slot allocation and hence transmission of its own data. Figure \ref{fig:aloha_qir} demonstrates the steps of updating the Q-values over three frames of a node that is allowed to transmit a maximum of two packets in each frame. Initially, the Q-values are initialized to zero, i.e., $Q\left(frame\#0\right)=\left\{ 0,0,0\right\}$. Upon successful transmission, the Q-value of each time slot is updated using the Q-learning update rule given by Eq. (\ref{eq:rl_update}) where the learning rate is set to 0.1. Upon successful transmission, the reward value is equal to +1, and it is -1 for a failed transmission. Certainly, the nodes will decide to transmit data using the time slots with the maximum Q-values. 

Although the idea of using reinforcement learning for duty cycle management is appealing because of its distributed operation and small memory and computational resource requirement, it may result in high collision rates during the initial exploration phases.

\subsubsection{Adaptive MAC layer}

\begin{figure}[t]
\begin{centering}
\includegraphics[trim=1cm 0.5cm 1cm 0.4cm, width=0.8\columnwidth]{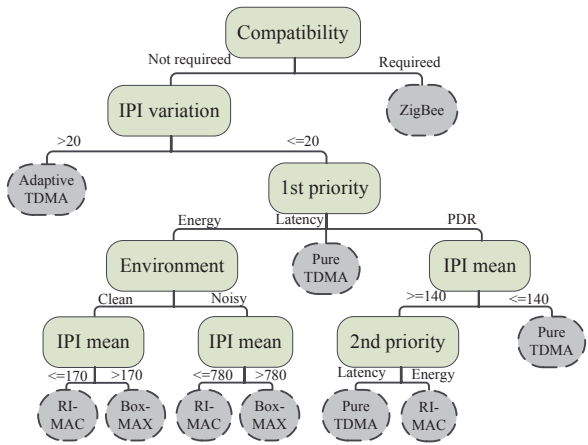}
\par\end{centering}

\caption{\label{fig:saml_dt} The decision tree classifier used to select the optimal MAC algorithm in the SAML architecture \cite{sha2013self}.}
\end{figure}

In a variety of modern applications, such as in healthcare and assisted living systems, WSNs are used to directly share the collected data with the users' mobile phones. This introduces new design challenges that are related to the dynamic communication patterns and service requirements over time. Sha \textit{et al.} \cite{sha2013self} studied this problem at the MAC layer, hence proposing the ``Self-Adapting MAC Layer'' (SAML) design. SAML is composed of two main components: The ``Reconfigurable MAC Architecture'' (RMA) to switch between the different MAC protocols, and the MAC engine that is used to learn the suitable MAC protocol for the current network conditions. The learning process is performed using the decision tree classier as illustrated in Fig. \ref{fig:saml_dt}. The learning features of the decision tree are: inter packet interval (IPI) and received signal strength indication (RSSI) statistical parameters (i.e., the mean and the variance), the application QoS requirements (reliability, energy usage, and latency), packet delivery rate (PDR), and the traffic pattern. The supported MAC protocols are Pure TDMA \cite{klues2007component}, Adaptive TDMA \cite{doerr2005multimac}, Box-MAC \cite{moss2008box}, RI-MAC \cite{sun2008ri}, and ZigBee \cite{alliance2007zigbee}. Even though the SAML scheme provides an adaptive MAC solution in dynamic environments, it introduces a level of complexity and additional expense into the designed systems.

\section{\label{sec:non_functional_challenges}NON-FUNCTIONAL CHALLENGES}

Non-functional requirements include specifications that are not related to the basic operational behavior of the system. For example, WSN designers may need to ensure that the proposed solution is always capable of providing up-to-date information about the monitored environment. This section provides a comprehensive review of recent machine learning advances that have been adopted to achieve non-functional requirements in WSNs such as security, quality of service, and data integrity. Moreover, this section also highlights some unique efforts in specialized WSN applications. Such studies could inspire researchers to a variety of WSN applications that can be improved using machine learning techniques.

\subsection{\label{sub:security_anomaly}Security and Anomaly Intrusion Detection}

\begin{figure}[t]
\begin{centering}
\includegraphics[trim=1.2cm 0.8cm 1.2cm 0.8cm, width=0.85\columnwidth]{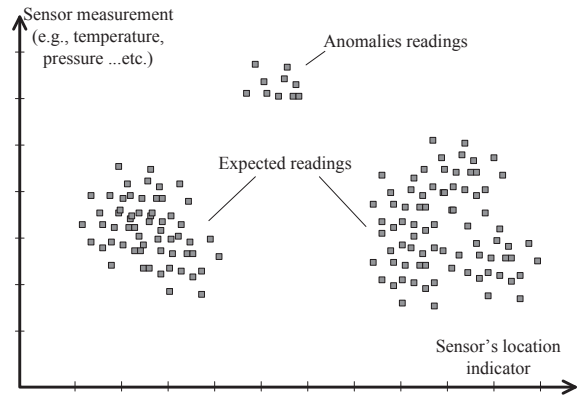}
\par\end{centering}

\caption{\label{fig:simple-ex-anomalies}An example of anomaly detection in phenomena monitoring sensor system using machine learning clustering and classification techniques (data set in euclidean space).}
\end{figure}

The major challenge to implement security techniques in WSNs is the limited resource constraints \cite{5451757}. Moreover, some attack methods aim to produce unexpected, mistaken knowledge, by introducing misleading observations to the network. Figure \ref{fig:simple-ex-anomalies} presents the general concept of anomaly detection in phenomenon monitoring sensor system using machine learning clustering and classification algorithms. In this example, machine learning techniques classify the data into two correct reading regions. Since most observations lie in these two regions, the points that are inconsistent (e.g., from an attack) with these regions are considered as anomalies.

Machine learning algorithms have been employed to detect outlying and misleading measurements. Simultaneously, several attacks could be detected by analyzing well-known malicious activities and vulnerabilities. Basically, WSN security enhancements by adopting machine learning techniques will result in the following earnings: 
\begin{itemize}
\item Save node's energy and significantly expand WSN lifetime by preventing the transmission of the outlier, misleading data.
\item Enhance network reliability by eliminating faulty and malicious readings. In the same way, avoiding the discovery of unexpected knowledge that will be converted to important, and often critical actions.
\item Online learning and prevention (without human intervention) of malicious attacks and vulnerabilities.
\end{itemize}
In this subsection, we explore various machine learning-based algorithms addressing the security issue in WSNs. Table \ref{tab:comparison-ml-outlier} summarizes the reviewed methods in this subsection. The column ``Predicting missing data'' indicates the ability of the proposed solution to provide predictions for any missing sensors' readings.

\begin{table*}[t]
\begin{centering}
\caption{\label{tab:comparison-ml-outlier}Summary of wireless sensor network outlier detection techniques that adopt machine learning paradigms.}

\begin{tabular}{|>{\centering}m{0.22\textwidth}|>{\centering}m{0.12\textwidth}|>{\centering}m{0.08\textwidth}|>{\centering}m{0.1\textwidth}|>{\centering}m{0.1\textwidth}|>{\centering}m{0.23\textwidth}|}
\hline 
\textbf{\noun{Studies}} & \textbf{\noun{Machine learning algorithm(s)}} & \textbf{\noun{Predicting missing data}} & \textbf{\noun{Distributed / centralized}} & \textbf{\noun{Complexity}} & \textbf{\noun{Aim(s)}}\tabularnewline
\hline 
\hline 
Outlier detection using Bayesian belief networks \cite{1665221} & Bayesian & Yes & Distributed & Low & Outlier detection\tabularnewline
\hline 
Outlier detection using k-NN \cite{1648838} & k-NN & Yes & Distributed & Moderate & Distributed outlier detection\tabularnewline
\hline 
Detecting selective forwarding attacks using SVM \cite{4496866} & \multirow{5}{0.08\textwidth}{\quad{}\enskip{}SVM} & \multirow{5}{0.08\textwidth}{\quad{}\quad{}No} & Centralized & Moderate & Detect black hole and selective forwarding attacks\tabularnewline
\cline{1-1} \cline{4-6} 
Distributed outlier detection using SVM \cite{4289308} &  &  & Distributed & Low & Outlier detection\tabularnewline
\cline{1-1} \cline{4-6} 
Online outlier detection \cite{4761978} &  &  & Centralized & Moderate & Online outlier detection\tabularnewline
\cline{1-1} \cline{4-6} 
Intrusion detection system \cite{2563258} &  &  & Centralized & High & Intrusion detection using immune and SVM algorithms\tabularnewline
\cline{1-1} \cline{4-6} 
Linear outlier detection \cite{Zhang20131062} &  &  & Distributed & Moderate & Adaptive outlier detection\tabularnewline
\hline 
Analyzing attacks with SOM \cite{4215510} & SOM & No & Distributed & Moderate & Detect anomalous behaviors\tabularnewline
\hline 
\end{tabular}\\
\par\end{centering}
\end{table*}

\subsubsection{Outlier detection using Bayesian belief network}

Janakiram \textit{et al.} \cite{1665221} used Bayesian belief networks (BBNs) to develop an outlier detection scheme. Given that the majority of node's neighbors will have similar readings (i.e., temporal and
spatial correlations), it is reasonable to use this phenomenon to build conditional dependencies among nodes' readings. BBNs infer the conditional relationships among the observations to discover any potential outliers in the collected data. Furthermore, this method can be used to evaluate missing values.

\subsubsection{Outlier detection using k-nearest neighbors}

Branch \textit{et al.} \cite{1648838} developed an in-network outlier detection method in WSNs using k-nearest neighbors. Moreover, any missing nodes' readings will be replaced by the average value of the k-nearest nodes. However, such non-parametric, k-NN-based algorithm requires large memory to store every collected readings from the monitored environment.

\subsubsection{Detecting selective forwarding attacks using support vector machine (SVM)}

In black hole attacks, malicious nodes send misleading ``Routing Reply'' (RREP) messages whenever the malicious nodes receive ``Route Request'' (RREQ) messages, indicating that routes to the destinations are found. Accordingly, source nodes will stop the process of route discovery, and will ignore other RREP messages. Therefore, malicious nodes will drop all network's messages, while the source nodes assume that their packets were delivered to the destination. Kaplantzis \textit{et al.} \cite{4496866} presented packet dropping attack prevention technique based on one class support vector machine classifier. The proposed scheme is capable of detecting black hole attacks and selective forwarding attacks. Basically, routing information, bandwidth and hop count are used to determine the malicious nodes in the network.

\subsubsection{Outlier detection using support vector machine (SVM)}

By using a quarter-sphere centered at the origin, the drawback of high computational requirements of traditional SVM could be alleviated. For instance, Rajasegarar \textit{et al.} \cite{4289308} introduced
a one-class quarter-sphere SVM anomaly recognition technique. The motivation of this distributed scheme is to distinguish anomalies in data while minimizing communication overhead. In \cite{4761978}, Yang \textit{et al.} tackled the design of an online outlier detection method using quarter-sphere SVM. The unsupervised learning method investigates the local data to reduce the computational complexity of traditional SVM-based outlier detection algorithms. This outlier detector is similar to the method introduced in \cite{4289308}.

Artificial immunity algorithm is a computationally intelligent algorithm for problem solving inspired by the biological immunity systems \cite{de2002artificial}. The biological immunity systems automatically generate the immune body (antibody) against the antigen (e.g., a virus) through the cell fission. In \cite{2563258}, Chen \textit{et al.} extended the basic idea of using SVM for detecting intrusion by combining it with immunity algorithm. In summary, an immune algorithm was introduced as a preprocessing step for the sensor data, that will be used by SVM to detect intruders. Furthermore, Zhang \textit{et al.} \cite{Zhang20131062} also investigated the temporal and spatial correlations of the collected readings using a one-class SVM learning algorithm to develop an outlier detection method. This study adopts an ellipsoidal one-class SVM that can be solved using linear optimization instead of the quadratic optimization problem in traditional SVM methods.

The main advantages of these SVM-based methods are their good performance (efficient learning) and ability to learn non-linear and complex problems. However, they still suffer from a scalability issue to large
data set due to their high computational and large memory requirements \cite{steinwart2008support}.

\subsubsection{Analyzing attacks with self-organizing map (SOM)}

Avram \textit{et al.} \cite{4215510} addressed the issue of detecting network attacks in wireless ad hoc networks using self-organizing map unsupervised learning. Learning the weights are obtained through statistical analysis of the input data vectors. The main issue of this scheme is the complexity in determining input weights. Moreover, SOM-based algorithms are not suitable for detecting attacks in very
large and complex data sets (i.e., large scale sensor network).

\subsection{Quality of Service, Data Integrity and Fault Detection}

Quality of service (QoS) guarantees high-priority delivery of real-time events and data. In the context of WSNs, there are potential multi-hop transmissions of data to the end user, in addition to distributing queries from  a system controller to the network nodes~\cite{7845239}. WSNs suffer from energy and bandwidth constraints that limit the quantity of information to be transmitted from a source to destination nodes. Furthermore, data aggregation and dissemination in WSNs can be faulty and unreliable \cite{451250}. These issues coupled with random network topologies introduce an important challenge for designing reliable algorithms for such networks. The state of the art and general QoS requirements in WSNs have been reviewed in \cite{chen2004qos}.

In the following, we review the latest efforts of using machine learning techniques to achieve specific QoS and data integrity constraints. In brief, this adoption results in the following advantages:
\begin{itemize}
\item Different machine learning classifiers are used to recognize different types of streams, thus eliminating the need for flow-aware management techniques.
\item The requirements for QoS guarantee, data integrity and fault detection depend on the network service and application. Machine learning methods are able to handle much of this while ensuring efficient resource utilization,
mainly bandwidth and power utilization.
\end{itemize}
Table \ref{tab:comparison_qos} summarizes the methods that are reviewed in this subsection. The column ``Characteristics'' indicates features or qualities belonging to each study.

\begin{table*}[t]
\begin{centering}
\caption{\label{tab:comparison_qos}Summary of quality of service, data integrity and fault detection solutions. }

\begin{tabular}{|>{\centering}m{0.28\textwidth}|>{\centering}m{0.12\textwidth}|>{\centering}m{0.1\textwidth}|>{\centering}m{0.4\textwidth}|}
\hline 
\textbf{\noun{Approaches}} & \textbf{\noun{Machine learning algorithm(s)}} & \textbf{\noun{Complexity}} & \textbf{\noun{Characteristics}}\tabularnewline
\hline 
\hline 
System's dependability \cite{1606090} & \multirow{2}{0.1\textwidth}{\quad{}\quad{}\enskip{}NNs} & High & Estimate the dependability metric\tabularnewline
\cline{1-1} \cline{3-4} 
Fault detection \cite{4429838} &  & Moderate & Dynamic fault detection model\tabularnewline
\hline 
MetricMap \cite{Wang:2007:PLQ:1317425.1317434} & \multirow{1}{0.1\textwidth}{\quad{}\quad{}\enskip{}DT} & Low & Link quality estimation\tabularnewline
\hline 
Assessing accuracy and reliability metrics \cite{Osborne:2008:TRI:1371607.1372727} & GP & Moderate & Information processing tasks\tabularnewline
\hline 
A QoS scheduler \cite{4230960} & \multirow{4}{0.1\textwidth}{\quad{}\quad{}\enskip{}RL} & \multicolumn{1}{c|}{Low} & QoS task scheduler for adaptive multimedia sensor networks\tabularnewline
\cline{1-1} \cline{3-4} 
Uncertainty and coverage factors \cite{4496881}  &  & Moderate & Investigate converge problems\tabularnewline
\cline{1-1} \cline{3-4} 
QoS-aware power management \cite{5284227} &  & Low & QoS-aware power management in energy harvesting  sensor nodes\tabularnewline
\cline{1-1} \cline{3-4} 
QoS provisioning \cite{4976244} &  & Low & A structure modeling tool for QoS provisioning\tabularnewline
\hline 
\end{tabular}
\par\end{centering}
\end{table*}

\subsubsection{QoS estimation using neural network}

Recently, there is growing interest in estimating and improving the performance of WSNs. For example, Snow \textit{et al.} \cite{1606090} introduced a method to estimate a sensor network dependability metric
using a neural network method. Dependability is a metric that represents availability, reliability, maintainability, and survivability of a sensor network. Several attributes are used to estimate such a metric  including mean time between failure (MTBF) and mean time to repair (MTTR).

Moustapha and Selmic \cite{4429838} introduced a dynamic fault detection model for WSNs. This model captures the nodes' dynamic behavior and their effects on other nodes. In addition, neural network learning,
which is trained using back-propagation method, was used for node identification and fault detection (a similar idea as in \cite{1606090}). This study results in an effective nonlinear sensor model that suits applications with fault detection requirements.

\subsubsection{MetricMap (link quality estimation framework)}

Link quality measurement tools may provide inaccurate and unstable readings across different environments due to different conditions such as signal variations and interference \cite{Baccour:2012:RLQ:2240116.2240123}. As a result, Wang \textit{et al.} \cite{Wang:2007:PLQ:1317425.1317434} presented MetricMap, a link quality estimation framework using supervised learning methods. MetricMap enhances the MintRoute \cite{Woo:2003:TUC:958491.958494} protocol by adopting online and offline learning methods, such as decision tree learners, to derive link quality indicators. This framework uses several local features to build the classification tree such as the received signal strength indicator (RSSI), transmission buffer size, channel load, and the forward and backward probabilities. The forward probability $p_{f}(l)$ is defined as the ratio of the received to the total transmitted packets over the link $l$, whereas the backward probability $p_{b}(l)$ is calculated over the reverse path. The local features are preferred over the global one as they can be found without costly communications with far away nodes. Experiments reveal that up to three times improvement in data delivery rate over basic MintRoute method can be achieved.

\subsubsection{Assessing accuracy and reliability of sensor nodes using multi-output Gaussian processes}

Osborne \textit{et al.} \cite{Osborne:2008:TRI:1371607.1372727} presented a real-time algorithm to determine a set of nodes that are capable of handling information processing tasks such as assessing the accuracy of collected sensor readings and predicting the missing readings. This algorithm provides a probabilistic Gaussian process based iterative implementation that is trained to re-use previous experience (i.e., the historical data) and maintain a reasonable training data size. Yet, the posterior distribution of an observed environmental variable $x$ (e.g., sea surface temperature) is calculated using the generalized multivariate Gaussian distribution given by Eq. (\ref{eq:qp_distribution}).

\begin{equation}
p\left(x|\mu,K,I\right)\triangleq\frac{1}{\sqrt{\det2\pi K}}\exp\left(-\frac{1}{2}\left(x-\mu\right)^{T}K^{-1}\left(x-\mu\right)\right)\label{eq:qp_distribution}
\end{equation}
where $\mu$, $K$ are the prior mean and covariance of the variable $x$, respectively. Further, $I$ denotes the historical data (a sequence of time-stamped samples) that is updated online to consider the new  sequentially collected observations.

\subsubsection{QoS provisioning using reinforcement learning}

Ouferhat and Mellouk \cite{4230960} introduced a QoS task scheduler for adaptive multimedia sensor networks based on Q-learning technique. This scheduler significantly enhances the network throughput by
reducing the transmission delay. Comparatively, Seah \textit{et al.} \cite{4496881} considered coverage as a QoS metric in WSNs that represents how efficiently the area of interest will be observed. A Q-learning method was used to develop a distributed learner that is able to find weakly monitored sites. These sites can be resolved in future re-deployment stages.

It is important to note that energy harvesting has not been considered in the above QoS mechanisms. Conversely, Hsu \textit{et al.} \cite{5284227} introduced a QoS-aware power management scheme for WSNs with energy harvesting capabilities, namely ``Reinforcement Learning based QoS-aware Power Management'' (RLPM). This scheme is able to adapt to the dynamic levels of nodes' energy in systems with energy harvesting capabilities. QoS-aware RLPM employs reinforcement learning to attain QoS awareness and to manage nodes' duty cycle under the energy restriction. Furthermore, Liang \textit{et al.} \cite{4976244} designed ``Multi-agent Reinforcement Learning based multi-hop mesh Cooperative Communication'' (MRL-CC) to be a structure modeling tool for QoS provisioning in WSNs. Basically, MRL-CC is adopted to reliably assess the data in a cooperative manner. Moreover, MRL-CC might be used to examine the impact of traffic load and node mobility on the whole network performance.

\subsection{Miscellaneous Applications}

This subsection presents miscellaneous and unique research efforts that are not discussed previously.

\subsubsection{Resource management through reinforcement learning}

\begin{figure}[t]
\begin{centering}
\includegraphics[trim=2.5cm 1.5cm 1cm 0.5cm, width=0.85\columnwidth]{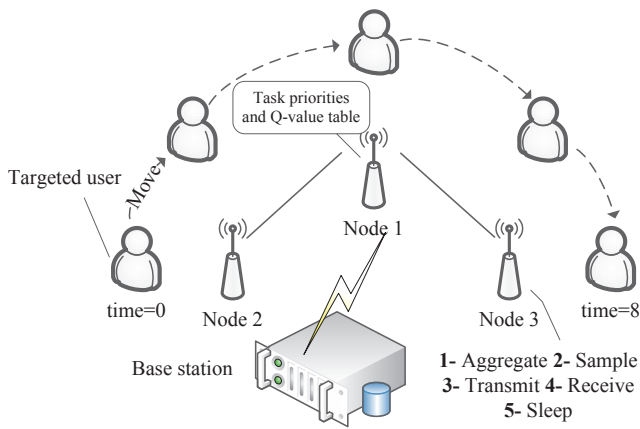}
\par\end{centering}

\caption{\label{fig:dirl}An example of task management using the DIRL middleware algorithm: Object tracking application \cite{4428658}.}
\end{figure}

Shah and Kumar \cite{4428658} presented the ``Distributed Independent Reinforcement Learning'' (DIRL) algorithm that utilizes local information and application constraints to optimize various tasks over time while minimizing energy consumption of the network. Each sensor node learns the minimum required resources to perform its scheduled tasks, and maximizes its future rewards by finding optimal parameters of the intended application. As a typical case, consider an object recognition application, as shown in Fig. \ref{fig:dirl}, which consists of five fundamental tasks: (a) aggregate two or more readings into a single reading, (b) transmit a message to the next hop, (c) receive incoming messages, (d) sample and take readings, and (e) put the radio into sleep mode. These tasks must be executed in some priority to maximize the network lifetime, where the network do not have such knowledge of priority as there is no static schedule for the events. For example, a node does not have the knowledge of when the object is going to move near to it to start taking samples. Here, the DIRL algorithm can be used to generate the required knowledge of priority by using the Q-learning algorithm and after specifying the set of reward and price value for each task.

\subsubsection{Decision tree-based animals behavior classification}

WSNs were applied in many applications such as environmental and habitat monitoring \cite{Mainwaring:2002:WSN:570738.570751}. As an example, Nadimi \textit{et al.} \cite{Nadimi2008167} employed decision tree to accurately classify the behavior of a herd of animals (active or inactive) using parameters such as the pitch angle of the neck and movement velocity. The advantages of the proposed solution for animals
behavior classification are the simple implementation and low complexity due to the use of a few critical features.

\subsubsection{Clock synchronization using self-organizing map}

Clock synchronization between sensor nodes is an important process, since most operations of the nodes must be consistent with each other. Moreover, the design of such methods for WSNs has to consider the limited resource constraints. For example, Paladina \textit{et al.} \cite{4689144} proposed to use SOM to ensure reliable clock synchronization for large scale networks. Nodes predict near-optimal estimation of the current time without having a central timing device and with limited storage and computing resources. However, this method assumes a uniform deployment of the nodes over the monitored area, as well as the same transmission powers for all nodes.

\subsubsection{Air quality monitoring using neural networks}

Postolache \textit{et al.} \cite{5196704} proposed a neural networks-based method for measuring air pollution levels using inexpensive gas sensor nodes, while eliminating the effects of temperature and humidity on sensor readings. This solution detects the air quality and gas concentration using neural networks implemented using JavaScript (JS). As a result, the solution is able to distribute processing between web server and end user computers (i.e., a combination of client and server side scripts).

\subsubsection{Intelligent lighting control using neural networks}

Gao \textit{et al.} \cite{gao2013wireless} introduced a new standard for lighting control in smart building using the neural network algorithm. A radial basis function (RBF) neural network is used to extract a new mathematical expression, called ``Illuminance Matrix'' (I-matrix), to measure the degree of illuminance in the lighted area. Fundamentally, in the field of lighting control, converting the collected data from the photosensors to a form that is suitable for digital signal processing is a crucial issue and can highly affect the performance of the developed system. The article shows that using the I-matrix scheme can achieve about 60\% more accuracy compared to the standard methods.

\section{\label{sec:open_issues}FUTURE APPLICATIONS OF MACHINE LEARNING IN WIRELESS SENSOR NETWORKS}
Although machine learning techniques have been applied to many applications in WSNs, many issues are still open and need further research efforts. 

\subsection{Compressive Sensing and Sparse Coding}

In practice, a large number of sensor measurements are usually required to maintain a desired detection accuracy. This introduces several challenges to network designers such as network management and communication issues. Given that 80 percent of the nodes' energy is consumed while sending and receiving data \cite{1425113}, data compression and dimensionality reduction techniques can be used to reduce transmission and hence prolong the network lifetime.

Traditional data compression techniques may result in extra energy consumption due to their high computational and memory requirements. In \cite{Barr:2006:ELD:1151690.1151692}, Barr and Asanovi\'{c} studied
the tradeoff between energy consumption in data transmission and compression. This study approximates the efficiency threshold of data compression in WSNs to be 1 bit data reduction using 485-1267 ADD instructions. 

Even though compressive sensing can be recast as a linear program, it still not applicable for on-node compression. As a result, it is important to apply and extend the basic concept of compressive sensing
to meet the resource constraint of WSNs. For more on the theoretical performance of decentralized compressive sensing, please refer to \cite{4472248,5502565,5707037}. Examples of similar emerging techniques
include independent component analysis, dictionary learning, non-negative matrix factorization and singular value decomposition.

\subsection{Distributed and Adaptive Machine Learning Techniques for WSNs}

Distributed machine learning techniques suit limited resource devices such as WSNs. Compared to centralized learning algorithms, distributed learning methods require less computational power and smaller
memory footprint (i.e., they do not need to consider the whole network information). The decentralized learning techniques enable the nodes to rapidly adapt their future behavior and predictions in tune with
the current environment conditions. For such reasons, distributed and adaptive learning algorithms are adequate for in-network processing of data while avoiding exhausting the nodes with high computational
tasks~\cite{7785419}. Examples of recent online learning algorithms include ``Adaptive Regularization of Weights'' (AROW) \cite{Koby:4525469}, ``Improved Ellipsoid Method for Online Learning'' (IELLIP) \cite{Yang:2009:OLE:1553374.1553521} and ``Soft Confidence-Weighted'' (SCW) \cite{wang2012exact}. Kotecha \textit{et al.} \cite{1413463} studied some distributed classification algorithms for WSNs.

\subsection{Resource Management Using Machine Learning}

Energy saving is a crucial issue in developing efficient WSNs algorithms and techniques. This design goal can be achieved using two main techniques, namely, by enhancing communication related protocols (e.g., routing and MAC protocols design) and by detecting nonfunctional and energy wasteful activities. The first technique includes physical, MAC and networking layer protocols. As it is discussed in this survey, this technique has been widely studied and enhanced using machine learning algorithms. The second technique focuses on decreasing the consumed energy in minor and nonfunctional requirements. For example, sensor nodes will consume their energy when over-listening to other nodes' transmissions \cite{jiang2007architecture}. Accordingly, such operations unnecessarily increase the active time of the nodes (i.e., increase nodes' duty cycle). The nodes that are equipped with machine learning techniques will be able to optimize their resource management and power allocation operations under those circumstances.

\subsection{Detecting Data Spatial and Temporal Correlations Using Hierarchical Clustering}

\begin{figure}[t]
\begin{centering}
\includegraphics[trim=1cm 1.2cm 1cm 1cm, width=0.75\columnwidth]{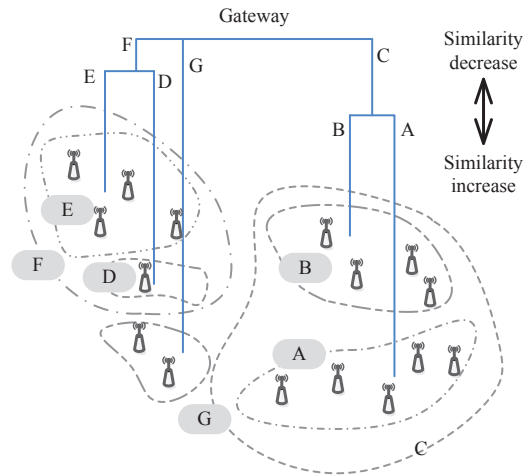}
\par\end{centering}

\caption{\label{fig:hierarchical_clustering}Hierarchical clustering of network's
nodes based on data spatial and temporal correlations in a temperature
monitoring system.}
\end{figure}

Hierarchical clustering is an unsupervised learning algorithm that aims to build a hierarchy of clusters. Basically, hierarchical clustering algorithms generate decomposition of the set of objects, which could
be a set of sensor nodes in WSNs. Broadly speaking, hierarchical clustering can provide an emerging clustering technique in WSNs using some clustering criteria such as spatial and temporal correlations of readings. Figure \ref{fig:hierarchical_clustering} illustrates such hierarchically clustered network based on spatial and temporal correlations of readings in a temperature monitoring system. In this example, Cluster C is formed by combining Clusters A and B, and so on for the rest of the clusters in the network.

The study of data correlation based on hierarchical clustering method will provide simple methods for energy saving. In such formations, only one node from each cluster is activated at a time to cover and monitor the whole cluster area. Typical methods of hierarchical clustering include ``Balanced Iterative Reducing and Clustering using Hierarchies'' (BIRCH) \cite{Zhang:1996:BED:235968.233324} and ``Clustering Using Representatives'' (CURE) \cite{Guha:1998:CEC:276304.276312}.

\section{\label{sec:CONCLUSIONS-AND-FUTURE}CONCLUSIONS}

Wireless sensor networks are different from traditional network in various aspects, thereby necessitating protocols and tools that address unique challenges and limitations. As a consequence, wireless sensor networks require innovative solutions for energy aware and real-time routing, security, scheduling, localization, node clustering, data aggregation, fault detection and data integrity. Machine learning provides a collection of techniques to enhance the ability of wireless sensor network to adapt to the dynamic behavior of its surrounding environment. Table \ref{tab:summary_publications} summarizes studies that have adopted machine learning methods to address these challenges from distinct research areas.

\begin{table*}[!t]
\begin{centering}
\caption{\label{tab:summary_publications}Summary of publications resolving various WSN challenges by the adoption of machine learning techniques.}

\begin{tabular}{|l||>{\centering}m{0.08\linewidth}|>{\centering}m{0.09\textwidth}|>{\centering}m{0.09\textwidth}|>{\centering}m{0.09\textwidth}|>{\centering}m{0.07\textwidth}||>{\centering}m{0.08\textwidth}|>{\centering}m{0.09\textwidth}|}
\cline{2-8}
\multicolumn{1}{l||}{\begin{sideways}
\end{sideways}} & \multicolumn{5}{c||}{\textbf{\noun{Functional}}} & \multicolumn{2}{c|}{\textbf{\noun{Non-functional}}}\tabularnewline
\cline{2-8} 
\multicolumn{1}{l||}{\begin{sideways}
\end{sideways}} & \textbf{\noun{\scriptsize{Routing}}} & \textbf{\noun{\scriptsize{Clustering and data aggregation }}} & \textbf{\noun{\scriptsize{Localization and objects targeting}}} & \textbf{\noun{\scriptsize{Event detection and query processing}}} & \textbf{\noun{MAC}} & \textbf{\noun{\scriptsize{Security and intrusion detection }}} & \textbf{\noun{\scriptsize{QoS, data integrity and fault detection }}}\tabularnewline
\hline 
\hline 
\textbf{\noun{Bayesian statistics }} &  &  & \cite{4518167,5161349} & \cite{1261832,1471678,5932548} & \cite{5612365} & \cite{1665221} & \tabularnewline
\hline 
\multicolumn{1}{|l||}{\textbf{\noun{k-nearest neighbors}}} &  &  &  & \cite{1541008,5698542} &  & \cite{1648838} & \tabularnewline
\hline 
\textbf{\noun{Neural network}} &  & \cite{4840489} & \cite{Shareef:2008:LUN:1361492.1361497,Yun:2009:SCA:1508324.1508595,6328975} & \cite{1544272} & \cite{Shen2008900,Kulkarni:2009:NNB:1704555.1704768} &  & \cite{1606090,4429838}\tabularnewline
\hline 
\textbf{\noun{Support vector machines}} &  &  & \cite{5506589,4384476,5487032} &  &  & \cite{4496866,4289308,4761978,2563258,Zhang20131062} & \tabularnewline
\hline 
\textbf{\noun{Decision tree }} &  & \cite{4761982} & \cite{4912210,Cayirci2006431} & \cite{5702151} & \cite{sha2013self} &  & \cite{Wang:2007:PLQ:1317425.1317434}\tabularnewline
\hline 
\textbf{\noun{Gaussian Processes}} &  & \cite{4301342,Kho:2009:DCA:1525856.1525857} & \cite{Krause:2008:NSP:1390681.1390689,6218781} &  &  &  & \cite{Osborne:2008:TRI:1371607.1372727}\tabularnewline
\hline 
\hline 
\textbf{\noun{Self-organizing map}} & \cite{9120642} & \cite{5489603} & \cite{4381576,4648079,Giorgetti:2007:WLU:1236360.1236399} &  &  & \cite{4215510} & \tabularnewline
\hline 
\textbf{\noun{Vector quantization}} &  & \cite{4024710} &  &  &  &  & \tabularnewline
\hline 
\textbf{\noun{Principal component}} &  & \cite{5425458,5345599,5706781,5671089,fenxiong2013data} &  & \cite{5486201} &  &  & \tabularnewline
\hline 
\textbf{\noun{k-means algorithm}} &  & \cite{4249813,985674} &  &  &  &  & \tabularnewline
\hline 
\hline 
\textbf{\noun{Reinforcement learning}} & \cite{1181362,4411037,4496872,4496810} & \cite{5158454,5486214} & \cite{6185486} & \cite{4428658} & \cite{Liu:2006:RRL:1359189.1359190,6328420} &  & \cite{4230960,4496881,5284227,4976244}\tabularnewline
\hline 
\end{tabular}
\end{centering}
\end{table*}

From the discussion so far, it became clear that many design challenges in wireless sensor networks have been resolved using several machine learning methods. In this paper, an extensive literature review over the period 2002-2013 on such studies was presented. In summary, adopting machine learning algorithms in wireless sensor networks has to consider the limited resources of the network, as well as the diversity of learning themes and patterns that will suit the problem at hand. Moreover, numerous issues are still open and need further research efforts such as developing lightweight and distributed message passing techniques, online learning algorithms, hierarchical clustering patterns and adopting machine learning in resource management problem of wireless sensor networks.

\bibliographystyle{IEEEtran}

\addcontentsline{toc}{section}{\refname}\bibliography{bibfiles/General_readings_wsn_challenges,bibfiles/Bayesian,bibfiles/mac_examples,bibfiles/General_readings_ml,bibfiles/summary,bibfiles/miscellaneous,bibfiles/Decision_tree,bibfiles/Gaussian_process_regression,bibfiles/K_means_algorithm,bibfiles/K_nearest_neighbor,bibfiles/Learning_vector_quantization,bibfiles/Neural_network,bibfiles/Principal_component,bibfiles/Reinforcement_learning,bibfiles/Self_organizing_map,bibfiles/Support_vector_machine}

\begin{thebibliography}{100}
\providecommand{\url}[1]{#1}
\csname url@samestyle\endcsname
\providecommand{\newblock}{\relax}
\providecommand{\bibinfo}[2]{#2}
\providecommand{\BIBentrySTDinterwordspacing}{\spaceskip=0pt\relax}
\providecommand{\BIBentryALTinterwordstretchfactor}{4}
\providecommand{\BIBentryALTinterwordspacing}{\spaceskip=\fontdimen2\font plus
\BIBentryALTinterwordstretchfactor\fontdimen3\font minus
  \fontdimen4\font\relax}
\providecommand{\BIBforeignlanguage}[2]{{%
\expandafter\ifx\csname l@#1\endcsname\relax
\typeout{** WARNING: IEEEtran.bst: No hyphenation pattern has been}%
\typeout{** loaded for the language `#1'. Using the pattern for}%
\typeout{** the default language instead.}%
\else
\language=\csname l@#1\endcsname
\fi
#2}}
\providecommand{\BIBdecl}{\relax}
\BIBdecl

\bibitem{ayodeleintroduction2010}
T.~O. Ayodele, ``Introduction to machine learning,'' in \emph{New Advances in
  Machine Learning}.\hskip 1em plus 0.5em minus 0.4em\relax InTech, 2010.

\bibitem{590079}
A.~H. Duffy, ``The ``what'' and ``how'' of learning in design,'' \emph{IEEE
  Expert}, vol.~12, no.~3, pp. 71--76, 1997.

\bibitem{Langley:1995:AML:219717.219768}
P.~Langley and H.~A. Simon, ``Applications of machine learning and rule
  induction,'' \emph{Communications of the ACM}, vol.~38, no.~11, pp. 54--64,
  1995.

\bibitem{451250}
L.~Paradis and Q.~Han, ``A survey of fault management in wireless sensor
  networks,'' \emph{Journal of Network and Systems Management}, vol.~15, no.~2,
  pp. 171--190, 2007.

\bibitem{1030829}
B.~Krishnamachari, D.~Estrin, and S.~Wicker, ``The impact of data aggregation
  in wireless sensor networks,'' in \emph{22nd International Conference on
  Distributed Computing Systems Workshops}, 2002, pp. 575--578.

\bibitem{1368893}
J.~Al-Karaki and A.~Kamal, ``Routing techniques in wireless sensor networks: A
  survey,'' \emph{IEEE Wireless Communications}, vol.~11, no.~6, pp. 6--28,
  2004.

\bibitem{romer2004design}
K.~Romer and F.~Mattern, ``The design space of wireless sensor networks,''
  \emph{IEEE Wireless Communications}, vol.~11, no.~6, pp. 54--61, 2004.

\bibitem{wan2013machine}
J.~Wan, M.~Chen, F.~Xia, L.~Di, and K.~Zhou, ``From machine-to-machine
  communications towards cyber-physical systems,'' \emph{Computer Science and
  Information Systems}, vol.~10, pp. 1105--1128, 2013.

\bibitem{bengio2009learning}
Y.~Bengio, ``Learning deep architectures for {AI},'' \emph{Foundations and
  Trends in Machine Learning}, vol.~2, no.~1, pp. 1--127, 2009.

\bibitem{hoffmann1990general}
A.~G. Hoffmann, ``General limitations on machine learning,'' pp. 345--347,
  1990.

\bibitem{4449882}
M.~Di and E.~M. Joo, ``A survey of machine learning in wireless sensor
  netoworks from networking and application perspectives,'' in \emph{6th
  International Conference on Information, Communications Signal Processing},
  2007, pp. 1--5.

\bibitem{4496871}
A.~Forster, ``Machine learning techniques applied to wireless ad-hoc networks:
  Guide and survey,'' in \emph{3rd International Conference on Intelligent
  Sensors, Sensor Networks and Information}.\hskip 1em plus 0.5em minus
  0.4em\relax IEEE, 2007, pp. 365--370.

\bibitem{forster2010machine}
A.~F{\"o}rster and M.~Amy~L, \emph{Machine learning across the {WSN}
  layers}.\hskip 1em plus 0.5em minus 0.4em\relax InTech, 2011.

\bibitem{5451757}
Y.~Zhang, N.~Meratnia, and P.~Havinga, ``Outlier detection techniques for
  wireless sensor networks: A survey,'' \emph{IEEE Communications Surveys \&
  Tutorials}, vol.~12, no.~2, pp. 159--170, 2010.

\bibitem{4589621}
V.~J. Hodge and J.~Austin, ``A survey of outlier detection methodologies,''
  \emph{Artificial Intelligence Review}, vol.~22, no.~2, pp. 85--126, 2004.

\bibitem{5473889}
R.~Kulkarni, A.~F{\"o}rster, and G.~Venayagamoorthy, ``Computational
  intelligence in wireless sensor networks: A survey,'' \emph{IEEE
  Communications Surveys \& Tutorials}, vol.~13, no.~1, pp. 68--96, 2011.

\bibitem{74851240}
S.~Das, A.~Abraham, and B.~K. Panigrahi, \emph{Computational intelligence:
  Foundations, perspectives, and recent trends}.\hskip 1em plus 0.5em minus
  0.4em\relax John Wiley \& Sons, Inc., 2010, pp. 1--37.

\bibitem{Abu-Mostafa:2012:LD:2207825}
Y.~S. Abu-Mostafa, M.~Magdon-Ismail, and H.-T. Lin, \emph{Learning from
  data}.\hskip 1em plus 0.5em minus 0.4em\relax AMLBook, 2012.

\bibitem{chapelle2006semi}
O.~Chapelle, B.~Schlkopf, and A.~Zien, \emph{Semi-supervised learning}.\hskip
  1em plus 0.5em minus 0.4em\relax MIT press Cambridge, 2006, vol.~2.

\bibitem{720536}
S.~Kulkarni, G.~Lugosi, and S.~Venkatesh, ``Learning pattern classification-a
  survey,'' \emph{IEEE Transactions on Information Theory}, vol.~44, no.~6, pp.
  2178--2206, 1998.

\bibitem{4518167}
M.~Morelande, B.~Moran, and M.~Brazil, ``Bayesian node localisation in wireless
  sensor networks,'' in \emph{IEEE International Conference on Acoustics,
  Speech and Signal Processing}, 2008, pp. 2545--2548.

\bibitem{5161349}
C.-H. Lu and L.-C. Fu, ``Robust location-aware activity recognition using
  wireless sensor network in an attentive home,'' \emph{IEEE Transactions on
  Automation Science and Engineering}, vol.~6, no.~4, pp. 598--609, 2009.

\bibitem{Shareef:2008:LUN:1361492.1361497}
A.~Shareef, Y.~Zhu, and M.~Musavi, ``Localization using neural networks in
  wireless sensor networks,'' in \emph{Proceedings of the 1st International
  Conference on Mobile Wireless Middleware, Operating Systems, and
  Applications}, 2008, pp. 1--7.

\bibitem{1541008}
J.~Winter, Y.~Xu, and W.-C. Lee, ``Energy efficient processing of k nearest
  neighbor queries in location-aware sensor networks,'' in \emph{2nd
  International Conference on Mobile and Ubiquitous Systems: Networking and
  Services}.\hskip 1em plus 0.5em minus 0.4em\relax IEEE, 2005, pp. 281--292.

\bibitem{5698542}
P.~P. Jayaraman, A.~Zaslavsky, and J.~Delsing, ``Intelligent processing of
  k-nearest neighbors queries using mobile data collectors in a location aware
  3{D} wireless sensor network,'' in \emph{Trends in Applied Intelligent
  Systems}.\hskip 1em plus 0.5em minus 0.4em\relax Springer, 2010, pp.
  260--270.

\bibitem{1544272}
L.~Yu, N.~Wang, and X.~Meng, ``Real-time forest fire detection with wireless
  sensor networks,'' in \emph{International Conference on Wireless
  Communications, Networking and Mobile Computing}, vol.~2, 2005, pp.
  1214--1217.

\bibitem{5702151}
M.~Bahrepour, N.~Meratnia, M.~Poel, Z.~Taghikhaki, and P.~J. Havinga,
  ``Distributed event detection in wireless sensor networks for disaster
  management,'' in \emph{2nd International Conference on Intelligent Networking
  and Collaborative Systems}.\hskip 1em plus 0.5em minus 0.4em\relax IEEE,
  2010, pp. 507--512.

\bibitem{5612365}
M.~Kim and M.-G. Park, ``Bayesian statistical modeling of system energy saving
  effectiveness for {MAC} protocols of wireless sensor networks,'' in
  \emph{Software Engineering, Artificial Intelligence, Networking and
  Parallel/Distributed Computing}, ser. Studies in Computational
  Intelligence.\hskip 1em plus 0.5em minus 0.4em\relax Springer Berlin
  Heidelberg, 2009, vol. 209, pp. 233--245.

\bibitem{Shen2008900}
Y.-J. Shen and M.-S. Wang, ``Broadcast scheduling in wireless sensor networks
  using fuzzy hopfield neural network,'' \emph{Expert Systems with
  Applications}, vol.~34, no.~2, pp. 900 -- 907, 2008.

\bibitem{Kulkarni:2009:NNB:1704555.1704768}
R.~V. Kulkarni and G.~K. Venayagamoorthy, ``Neural network based secure media
  access control protocol for wireless sensor networks,'' in \emph{Proceedings
  of the 2009 International Joint Conference on Neural Networks}, ser.
  IJCNN'09.\hskip 1em plus 0.5em minus 0.4em\relax Piscataway, NJ, USA: IEEE
  Press, 2009, pp. 3437--3444.

\bibitem{1665221}
D.~Janakiram, V.~Adi Mallikarjuna~Reddy, and A.~Phani~Kumar, ``Outlier
  detection in wireless sensor networks using {B}ayesian belief networks,'' in
  \emph{1st International Conference on Communication System Software and
  Middleware}.\hskip 1em plus 0.5em minus 0.4em\relax IEEE, 2006, pp. 1--6.

\bibitem{1648838}
J.~W. Branch, C.~Giannella, B.~Szymanski, R.~Wolff, and H.~Kargupta,
  ``In-network outlier detection in wireless sensor networks,'' \emph{Knowledge
  and information systems}, vol.~34, no.~1, pp. 23--54, 2013.

\bibitem{4496866}
S.~Kaplantzis, A.~Shilton, N.~Mani, and Y.~Sekercioglu, ``Detecting selective
  forwarding attacks in wireless sensor networks using support vector
  machines,'' in \emph{3rd International Conference on Intelligent Sensors,
  Sensor Networks and Information}.\hskip 1em plus 0.5em minus 0.4em\relax
  IEEE, 2007, pp. 335--340.

\bibitem{4289308}
S.~Rajasegarar, C.~Leckie, M.~Palaniswami, and J.~Bezdek, ``Quarter sphere
  based distributed anomaly detection in wireless sensor networks,'' in
  \emph{International Conference on Communications}, 2007, pp. 3864--3869.

\bibitem{1606090}
A.~Snow, P.~Rastogi, and G.~Weckman, ``Assessing dependability of wireless
  networks using neural networks,'' in \emph{Military Communications
  Conference}.\hskip 1em plus 0.5em minus 0.4em\relax IEEE, 2005, pp.
  2809--2815 Vol. 5.

\bibitem{4429838}
A.~Moustapha and R.~Selmic, ``Wireless sensor network modeling using modified
  recurrent neural networks: Application to fault detection,'' \emph{IEEE
  Transactions on Instrumentation and Measurement}, vol.~57, no.~5, pp.
  981--988, 2008.

\bibitem{Wang:2007:PLQ:1317425.1317434}
Y.~Wang, M.~Martonosi, and L.-S. Peh, ``Predicting link quality using
  supervised learning in wireless sensor networks,'' \emph{ACM SIGMOBILE Mobile
  Computing and Communications Review}, vol.~11, no.~3, pp. 71--83, 2007.

\bibitem{beyer1999nearest}
K.~Beyer, J.~Goldstein, R.~Ramakrishnan, and U.~Shaft, ``When is ``nearest
  neighbor'' meaningful?'' in \emph{Database Theory}.\hskip 1em plus 0.5em
  minus 0.4em\relax Springer, 1999, pp. 217--235.

\bibitem{ayodele2010types}
T.~O. Ayodele, ``Types of machine learning algorithms,'' in \emph{New Advances
  in Machine Learning}.\hskip 1em plus 0.5em minus 0.4em\relax InTech, 2010.

\bibitem{safavian1991survey}
S.~R. Safavian and D.~Landgrebe, ``A survey of decision tree classifier
  methodology,'' \emph{IEEE Transactions on Systems, Man and Cybernetics},
  vol.~21, no.~3, pp. 660--674, 1991.

\bibitem{lippmann1987introduction}
R.~Lippmann, ``An introduction to computing with neural nets,'' \emph{ASSP
  Magazine, IEEE}, vol.~4, no.~2, pp. 4--22, 1987.

\bibitem{8547966}
W.~Dargie and C.~Poellabauer, \emph{Localization}.\hskip 1em plus 0.5em minus
  0.4em\relax John Wiley \& Sons, Ltd, 2010, pp. 249--266.

\bibitem{Kohonen:2001}
T.~Kohonen, \emph{\BIBforeignlanguage{English}{Self-organizing maps}}, ser.
  Springer Series in Information Sciences.\hskip 1em plus 0.5em minus
  0.4em\relax Springer Berlin Heidelberg, 2001, vol.~30.

\bibitem{hinton2006reducing}
G.~E. Hinton and R.~R. Salakhutdinov, ``Reducing the dimensionality of data
  with neural networks,'' \emph{Science}, vol. 313, no. 5786, pp. 504--507,
  2006.

\bibitem{steinwart2008support}
I.~Steinwart and A.~Christmann, \emph{Support vector machines}.\hskip 1em plus
  0.5em minus 0.4em\relax Springer, 2008.

\bibitem{4761978}
Z.~Yang, N.~Meratnia, and P.~Havinga, ``An online outlier detection technique
  for wireless sensor networks using unsupervised quarter-sphere support vector
  machine,'' in \emph{International Conference on Intelligent Sensors, Sensor
  Networks and Information Processing}.\hskip 1em plus 0.5em minus 0.4em\relax
  IEEE, 2008, pp. 151--156.

\bibitem{2563258}
Y.~Chen, Y.~Qin, Y.~Xiang, J.~Zhong, and X.~Jiao, ``Intrusion detection system
  based on immune algorithm and support vector machine in wireless sensor
  network,'' in \emph{Information and Automation}, ser. Communications in
  Computer and Information Science.\hskip 1em plus 0.5em minus 0.4em\relax
  Springer Berlin Heidelberg, 2011, vol.~86, pp. 372--376.

\bibitem{Zhang20131062}
Y.~Zhang, N.~Meratnia, and P.~J. Havinga, ``Distributed online outlier
  detection in wireless sensor networks using ellipsoidal support vector
  machine,'' \emph{Ad Hoc Networks}, vol.~11, no.~3, pp. 1062--1074, 2013.

\bibitem{5506589}
W.~Kim, J.~Park, and H.~Kim, ``Target localization using ensemble support
  vector regression in wireless sensor networks,'' in \emph{Wireless
  Communications and Networking Conference}, 2010, pp. 1--5.

\bibitem{4384476}
D.~Tran and T.~Nguyen, ``Localization in wireless sensor networks based on
  support vector machines,'' \emph{IEEE Transactions on Parallel and
  Distributed Systems}, vol.~19, no.~7, pp. 981--994, 2008.

\bibitem{5487032}
B.~Yang, J.~Yang, J.~Xu, and D.~Yang, ``Area localization algorithm for mobile
  nodes in wireless sensor networks based on support vector machines,'' in
  \emph{Mobile Ad-Hoc and Sensor Networks}.\hskip 1em plus 0.5em minus
  0.4em\relax Springer, 2007, pp. 561--571.

\bibitem{box2011bayesian}
G.~E. Box and G.~C. Tiao, \emph{Bayesian inference in statistical
  analysis}.\hskip 1em plus 0.5em minus 0.4em\relax John Wiley \& Sons, 2011,
  vol.~40.

\bibitem{Rasmussen06gaussianprocesses}
C.~E. Rasmussen, ``Gaussian processes for machine learning,'' in \emph{in:
  Adaptive Computation and Machine Learning}.\hskip 1em plus 0.5em minus
  0.4em\relax Citeseer, 2006.

\bibitem{5489603}
S.~Lee and T.~Chung, ``Data aggregation for wireless sensor networks using
  self-organizing map,'' in \emph{Artificial Intelligence and Simulation}, ser.
  Lecture Notes in Computer Science.\hskip 1em plus 0.5em minus 0.4em\relax
  Springer Berlin Heidelberg, 2005, vol. 3397, pp. 508--517.

\bibitem{5425458}
R.~Masiero, G.~Quer, D.~Munaretto, M.~Rossi, J.~Widmer, and M.~Zorzi, ``Data
  acquisition through joint compressive sensing and principal component
  analysis,'' in \emph{Global Telecommunications Conference}.\hskip 1em plus
  0.5em minus 0.4em\relax IEEE, 2009, pp. 1--6.

\bibitem{5345599}
R.~Masiero, G.~Quer, M.~Rossi, and M.~Zorzi, ``A {B}ayesian analysis of
  compressive sensing data recovery in wireless sensor networks,'' in
  \emph{International Conference on Ultra Modern Telecommunications Workshops},
  2009, pp. 1--6.

\bibitem{5706781}
A.~Rooshenas, H.~Rabiee, A.~Movaghar, and M.~Naderi, ``Reducing the data
  transmission in wireless sensor networks using the principal component
  analysis,'' in \emph{6th International Conference on Intelligent Sensors,
  Sensor Networks and Information Processing}.\hskip 1em plus 0.5em minus
  0.4em\relax IEEE, 2010, pp. 133--138.

\bibitem{5671089}
S.~Macua, P.~Belanovic, and S.~Zazo, ``Consensus-based distributed principal
  component analysis in wireless sensor networks,'' in \emph{11th International
  Workshop on Signal Processing Advances in Wireless Communications}, 2010, pp.
  1--5.

\bibitem{4249813}
Y.-C. Tseng, Y.-C. Wang, K.-Y. Cheng, and Y.-Y. Hsieh, ``i{M}ouse: An
  integrated mobile surveillance and wireless sensor system,'' \emph{Computer},
  vol.~40, no.~6, pp. 60--66, 2007.

\bibitem{985674}
D.~Li, K.~Wong, Y.~H. Hu, and A.~Sayeed, ``Detection, classification, and
  tracking of targets,'' \emph{IEEE Signal Processing Magazine}, vol.~19,
  no.~2, pp. 17--29, 2002.

\bibitem{kanungo2002efficient}
T.~Kanungo, D.~M. Mount, N.~S. Netanyahu, C.~D. Piatko, R.~Silverman, and A.~Y.
  Wu, ``An efficient k-means clustering algorithm: Analysis and
  implementation,'' \emph{IEEE Transactions on Pattern Analysis and Machine
  Intelligence}, vol.~24, no.~7, pp. 881--892, 2002.

\bibitem{jolliffe2002principal}
I.~T. Jolliffe, \emph{Principal component analysis}.\hskip 1em plus 0.5em minus
  0.4em\relax Springer verlag, 2002.

\bibitem{feldman2013turning}
D.~Feldman, M.~Schmidt, C.~Sohler, D.~Feldman, M.~Schmidt, and C.~Sohler,
  ``Turning big data into tiny data: Constant-size coresets for k-means, {PCA}
  and projective clustering,'' in \emph{SODA}, 2013, pp. 1434--1453.

\bibitem{15852369}
C.~Watkins and P.~Dayan, ``\BIBforeignlanguage{English}{Q-learning},''
  \emph{\BIBforeignlanguage{English}{Machine Learning}}, vol.~8, no. 3-4, pp.
  279--292, 1992.

\bibitem{1181362}
R.~Sun, S.~Tatsumi, and G.~Zhao, ``{Q-MAP}: A novel multicast routing method in
  wireless ad hoc networks with multiagent reinforcement learning,'' in
  \emph{Region 10 Conference on Computers, Communications, Control and Power
  Engineering}, vol.~1, 2002, pp. 667--670 vol.1.

\bibitem{4411037}
S.~Dong, P.~Agrawal, and K.~Sivalingam, ``Reinforcement learning based
  geographic routing protocol for {UWB} wireless sensor network,'' in
  \emph{Global Telecommunications Conference}.\hskip 1em plus 0.5em minus
  0.4em\relax IEEE, 2007, pp. 652--656.

\bibitem{4496872}
A.~F{\"o}rster and A.~Murphy, ``{FROMS}: Feedback routing for optimizing
  multiple sinks in wsn with reinforcement learning,'' in \emph{3rd
  International Conference on Intelligent Sensors, Sensor Networks and
  Information}.\hskip 1em plus 0.5em minus 0.4em\relax IEEE, 2007, pp.
  371--376.

\bibitem{4496810}
R.~Arroyo-Valles, R.~Alaiz-Rodriguez, A.~Guerrero-Curieses, and J.~Cid-Sueiro,
  ``Q-probabilistic routing in wireless sensor networks,'' in \emph{3rd
  International Conference on Intelligent Sensors, Sensor Networks and
  Information}.\hskip 1em plus 0.5em minus 0.4em\relax IEEE, 2007, pp. 1--6.

\bibitem{1307317}
C.~Guestrin, P.~Bodik, R.~Thibaux, M.~Paskin, and S.~Madden, ``Distributed
  regression: An efficient framework for modeling sensor network data,'' in
  \emph{3rd International Symposium on Information Processing in Sensor
  Networks}, 2004, pp. 1--10.

\bibitem{9120642}
J.~Barbancho, C.~Le{\'o}n, F.~Molina, and A.~Barbancho,
  ``\BIBforeignlanguage{English}{A new {QoS} routing algorithm based on
  self-organizing maps for wireless sensor networks},''
  \emph{\BIBforeignlanguage{English}{Telecommunication Systems}}, vol.~36, pp.
  73--83, 2007.

\bibitem{Scholkopf:2001:LKS:559923}
B.~Scholkopf and A.~J. Smola, \emph{Learning with kernels: Support vector
  machines, regularization, optimization, and beyond}.\hskip 1em plus 0.5em
  minus 0.4em\relax Cambridge, MA, USA: MIT Press, 2001.

\bibitem{1315937}
J.~Kivinen, A.~Smola, and R.~Williamson, ``Online learning with kernels,''
  \emph{IEEE Transactions on Signal Processing}, vol.~52, no.~8, pp.
  2165--2176, 2004.

\bibitem{1201597}
G.~Aiello and G.~Rogerson, ``Ultra-wideband wireless systems,'' \emph{IEEE
  Microwave Magazine}, vol.~4, no.~2, pp. 36--47, 2003.

\bibitem{4062839}
R.~Rajagopalan and P.~Varshney, ``Data-aggregation techniques in sensor
  networks: A survey,'' \emph{IEEE Communications Surveys \& Tutorials},
  vol.~8, no.~4, pp. 48--63, 2006.

\bibitem{1630358}
G.~Crosby, N.~Pissinou, and J.~Gadze, ``A framework for trust-based cluster
  head election in wireless sensor networks,'' in \emph{2nd IEEE Workshop on
  Dependability and Security in Sensor Networks and Systems}, 2006, pp. 10--22.

\bibitem{4493846}
J.-M. Kim, S.-H. Park, Y.-J. Han, and T.-M. Chung, ``{CHEF}: Cluster head
  election mechanism using fuzzy logic in wireless sensor networks,'' in
  \emph{10th International Conference on Advanced Communication Technology},
  vol.~1.\hskip 1em plus 0.5em minus 0.4em\relax IEEE, 2008, pp. 654--659.

\bibitem{1420160}
S.~Soro and W.~Heinzelman, ``Prolonging the lifetime of wireless sensor
  networks via unequal clustering,'' in \emph{19th IEEE International Parallel
  and Distributed Processing Symposium}, 2005, pp. 4--8.

\bibitem{abbasi2007survey}
A.~A. Abbasi and M.~Younis, ``A survey on clustering algorithms for wireless
  sensor networks,'' \emph{Computer communications}, vol.~30, no.~14, pp.
  2826--2841, 2007.

\bibitem{4840489}
H.~He, Z.~Zhu, and E.~Makinen, ``A neural network model to minimize the
  connected dominating set for self-configuration of wireless sensor
  networks,'' \emph{IEEE Transactions on Neural Networks}, vol.~20, no.~6, pp.
  973--982, 2009.

\bibitem{4761982}
G.~Ahmed, N.~M. Khan, Z.~Khalid, and R.~Ramer, ``Cluster head selection using
  decision trees for wireless sensor networks,'' in \emph{International
  Conference on Intelligent Sensors, Sensor Networks and Information
  Processing}.\hskip 1em plus 0.5em minus 0.4em\relax IEEE, 2008, pp. 173--178.

\bibitem{4301342}
E.~Ertin, ``Gaussian process models for censored sensor readings,'' in
  \emph{14th Workshop on Statistical Signal Processing}.\hskip 1em plus 0.5em
  minus 0.4em\relax IEEE, 2007, pp. 665--669.

\bibitem{Kho:2009:DCA:1525856.1525857}
J.~Kho, A.~Rogers, and N.~R. Jennings, ``Decentralized control of adaptive
  sampling in wireless sensor networks,'' \emph{ACM Transactions on Sensor
  Networks (TOSN)}, vol.~5, no.~3, pp. 19:1--19:35, 2009.

\bibitem{4024710}
S.~Lin, V.~Kalogeraki, D.~Gunopulos, and S.~Lonardi, ``Online information
  compression in sensor networks,'' in \emph{IEEE International Conference on
  Communications}, vol.~7.\hskip 1em plus 0.5em minus 0.4em\relax IEEE, 2006,
  pp. 3371--3376.

\bibitem{fenxiong2013data}
C.~Fenxiong, L.~Mingming, W.~Dianhong, and T.~Bo, ``Data compression through
  principal component analysis over wireless sensor networks,'' \emph{Journal
  of Computational Information Systems}, vol.~9, no.~5, pp. 1809--1816, 2013.

\bibitem{5158454}
A.~F{\"o}rster and A.~Murphy, ``{CLIQUE}: Role-free clustering with q-learning
  for wireless sensor networks,'' in \emph{29th IEEE International Conference
  on Distributed Computing Systems}, 2009, pp. 441--449.

\bibitem{5486214}
M.~Mihaylov, K.~Tuyls, and A.~Nowe, ``Decentralized learning in wireless sensor
  networks,'' in \emph{Adaptive and Learning Agents}, ser. Lecture Notes in
  Computer Science.\hskip 1em plus 0.5em minus 0.4em\relax Springer Berlin
  Heidelberg, 2010, vol. 5924, pp. 60--73.

\bibitem{heinzelman2000application}
W.~B. Heinzelman, ``Application-specific protocol architectures for wireless
  networks,'' Ph.D. dissertation, Massachusetts Institute of Technology, 2000.

\bibitem{5954192}
M.~Duarte and Y.~Eldar, ``Structured compressed sensing: From theory to
  applications,'' \emph{IEEE Transactions on Signal Processing}, vol.~59,
  no.~9, pp. 4053--4085, 2011.

\bibitem{dempster1977maximum}
A.~P. Dempster, N.~M. Laird, and D.~B. Rubin, ``Maximum likelihood from
  incomplete data via the {EM} algorithm,'' \emph{Journal of the Royal
  Statistical Society. Series B (Methodological)}, pp. 1--38, 1977.

\bibitem{degroot1974reaching}
M.~H. DeGroot, ``Reaching a consensus,'' \emph{Journal of the American
  Statistical Association}, vol.~69, no. 345, pp. 118--121, 1974.

\bibitem{1261832}
B.~Krishnamachari and S.~Iyengar, ``Distributed bayesian algorithms for
  fault-tolerant event region detection in wireless sensor networks,''
  \emph{IEEE Transactions on Computers}, vol.~53, no.~3, pp. 241--250, 2004.

\bibitem{5932548}
P.~Zappi, C.~Lombriser, T.~Stiefmeier, E.~Farella, D.~Roggen, L.~Benini, and
  G.~Tr{\"o}ster, ``Activity recognition from on-body sensors: Accuracy-power
  trade-off by dynamic sensor selection,'' in \emph{Wireless Sensor
  Networks}.\hskip 1em plus 0.5em minus 0.4em\relax Springer, 2008, pp. 17--33.

\bibitem{5486201}
H.~Malik, A.~Malik, and C.~Roy, ``\BIBforeignlanguage{English}{A methodology to
  optimize query in wireless sensor networks using historical data},''
  \emph{\BIBforeignlanguage{English}{Journal of Ambient Intelligence and
  Humanized Computing}}, vol.~2, pp. 227--238, 2011.

\bibitem{1471678}
Q.~Chen, K.-Y. Lam, and P.~Fan, ``Comments on "{D}istributed {B}ayesian
  algorithms for fault-tolerant event region detection in wireless sensor
  networks",'' \emph{IEEE Transactions on Computers}, vol.~54, no.~9, pp.
  1182--1183, 2005.

\bibitem{4017702}
K.~Sha, W.~Shi, and O.~Watkins, ``Using wireless sensor networks for fire
  rescue applications: Requirements and challenges,'' in \emph{IEEE
  International Conference on Electro/information Technology}, 2006, pp.
  239--244.

\bibitem{4343996}
H.~Liu, H.~Darabi, P.~Banerjee, and J.~Liu, ``Survey of wireless indoor
  positioning techniques and systems,'' \emph{IEEE Transactions on Systems,
  Man, and Cybernetics, Part C: Applications and Reviews}, vol.~37, no.~6, pp.
  1067--1080, 2007.

\bibitem{56532056}
J.~Wang, R.~Ghosh, and S.~Das, ``\BIBforeignlanguage{English}{A survey on
  sensor localization},'' \emph{\BIBforeignlanguage{English}{Journal of Control
  Theory and Applications}}, vol.~8, no.~1, pp. 2--11, 2010.

\bibitem{Nasipuri:2002:DBL:570738.570754}
A.~Nasipuri and K.~Li, ``A directionality based location discovery scheme for
  wireless sensor networks,'' in \emph{Proceedings of the 1st ACM International
  Workshop on Wireless Sensor Networks and Applications}.\hskip 1em plus 0.5em
  minus 0.4em\relax ACM, 2002, pp. 105--111.

\bibitem{Yun:2009:SCA:1508324.1508595}
S.~Yun, J.~Lee, W.~Chung, E.~Kim, and S.~Kim, ``A soft computing approach to
  localization in wireless sensor networks,'' \emph{Expert Systems with
  Applications}, vol.~36, no.~4, pp. 7552--7561, 2009.

\bibitem{6328975}
S.~Chagas, J.~Martins, and L.~de~Oliveira, ``An approach to localization scheme
  of wireless sensor networks based on artificial neural networks and genetic
  algorithms,'' in \emph{10th International Conference on New Circuits and
  Systems}.\hskip 1em plus 0.5em minus 0.4em\relax IEEE, 2012, pp. 137--140.

\bibitem{4912210}
Z.~Merhi, M.~Elgamel, and M.~Bayoumi, ``A lightweight collaborative fault
  tolerant target localization system for wireless sensor networks,''
  \emph{IEEE Transactions on Mobile Computing}, vol.~8, no.~12, pp. 1690--1704,
  2009.

\bibitem{Cayirci2006431}
E.~Cayirci, H.~Tezcan, Y.~Dogan, and V.~Coskun, ``Wireless sensor networks for
  underwater survelliance systems,'' \emph{Ad Hoc Networks}, vol.~4, no.~4, pp.
  431--446, 2006.

\bibitem{Krause:2008:NSP:1390681.1390689}
A.~Krause, A.~Singh, and C.~Guestrin, ``Near-optimal sensor placements in
  gaussian processes: Theory, efficient algorithms and empirical studies,''
  \emph{The Journal of Machine Learning Research}, vol.~9, pp. 235--284, 2008.

\bibitem{6218781}
D.~Gu and H.~Hu, ``Spatial {G}aussian process regression with mobile sensor
  networks,'' \emph{IEEE Transactions on Neural Networks and Learning Systems},
  vol.~23, no.~8, pp. 1279--1290, 2012.

\bibitem{4381576}
L.~Paladina, M.~Paone, G.~Iellamo, and A.~Puliafito, ``Self organizing maps for
  distributed localization in wireless sensor networks,'' in \emph{12th IEEE
  Symposium on Computers and Communications}, 2007, pp. 1113--1118.

\bibitem{Giorgetti:2007:WLU:1236360.1236399}
G.~Giorgetti, S.~K.~S. Gupta, and G.~Manes, ``Wireless localization using
  self-organizing maps,'' in \emph{Proceedings of the 6th International
  Conference on Information Processing in Sensor Networks}, ser. IPSN
  '07.\hskip 1em plus 0.5em minus 0.4em\relax New York, NY, USA: ACM, 2007, pp.
  293--302.

\bibitem{4648079}
J.~Hu and G.~Lee, ``Distributed localization of wireless sensor networks using
  self-organizing maps,'' in \emph{IEEE International Conference on Multisensor
  Fusion and Integration for Intelligent Systems}, 2008, pp. 284--289.

\bibitem{6185486}
S.~Li, X.~Kong, and D.~Lowe, ``Dynamic path determination of mobile beacons
  employing reinforcement learning for wireless sensor localization,'' in
  \emph{26th International Conference on Advanced Information Networking and
  Applications Workshops}, 2012, pp. 760--765.

\bibitem{musso2001improving}
C.~Musso, N.~Oudjane, and F.~Le~Gland, ``Improving regularised particle
  filters,'' in \emph{Sequential Monte Carlo methods in practice}.\hskip 1em
  plus 0.5em minus 0.4em\relax Springer, 2001, pp. 247--271.

\bibitem{6165290}
Y.-X. Wang and Y.-J. Zhang, ``Non-negative matrix factorization: A
  comprehensive review,'' \emph{IEEE Transactions on Knowledge and Data
  Engineering}, vol.~25, no.~6, pp. 1336--1353, 2013.

\bibitem{tan2011survey}
H.-P. Tan, R.~Diamant, W.~K. Seah, and M.~Waldmeyer, ``A survey of techniques
  and challenges in underwater localization,'' \emph{Ocean Engineering},
  vol.~38, no.~14, pp. 1663--1676, 2011.

\bibitem{6328420}
Y.~Chu, P.~Mitchell, and D.~Grace, ``{ALOHA} and q-learning based medium access
  control for wireless sensor networks,'' in \emph{International Symposium on
  Wireless Communication Systems}, 2012, pp. 511--515.

\bibitem{5451759}
A.~Bachir, M.~Dohler, T.~Watteyne, and K.~K. Leung, ``{MAC} essentials for
  wireless sensor networks,'' \emph{IEEE Communications Surveys \& Tutorials},
  vol.~12, no.~2, pp. 222--248, 2010.

\bibitem{Liu:2006:RRL:1359189.1359190}
Z.~Liu and I.~Elhanany, ``{RL-MAC}: A reinforcement learning based {MAC}
  protocol for wireless sensor networks,'' \emph{International Journal of
  Sensor Networks}, vol.~1, no.~3, pp. 117--124, 2006.

\bibitem{sha2013self}
M.~Sha, R.~Dor, G.~Hackmann, C.~Lu, T.-S. Kim, and T.~Park, ``Self-adapting
  {MAC} layer for wireless sensor networks,'' Technical Report WUCSE-2013-75,
  Washington University in St. Louis, Tech. Rep., 2013.

\bibitem{1019408}
W.~Ye, J.~Heidemann, and D.~Estrin, ``An energy-efficient {MAC} protocol for
  wireless sensor networks,'' in \emph{21st Annual Joint Conference of the IEEE
  Computer and Communications Societies}, vol.~3, 2002, pp. 1567--1576 vol.3.

\bibitem{vanDam:2003:AEM:958491.958512}
T.~van Dam and K.~Langendoen, ``An adaptive energy-efficient {MAC} protocol for
  wireless sensor networks,'' in \emph{Proceedings of the 1st International
  Conference on Embedded Networked Sensor Systems}, ser. SenSys '03.\hskip 1em
  plus 0.5em minus 0.4em\relax New York, NY, USA: ACM, 2003, pp. 171--180.

\bibitem{klues2007component}
K.~Klues, G.~Hackmann, O.~Chipara, and C.~Lu, ``A component-based architecture
  for power-efficient media access control in wireless sensor networks,'' in
  \emph{Proceedings of the 5th International Conference on Embedded Networked
  Sensor Systems}.\hskip 1em plus 0.5em minus 0.4em\relax ACM, 2007, pp.
  59--72.

\bibitem{doerr2005multimac}
C.~Doerr, M.~Neufeld, J.~Fifield, T.~Weingart, D.~C. Sicker, and D.~Grunwald,
  ``{MultiMAC}-an adaptive {MAC} framework for dynamic radio networking,'' in
  \emph{International Symposium on New Frontiers in Dynamic Spectrum Access
  Networks}.\hskip 1em plus 0.5em minus 0.4em\relax IEEE, 2005, pp. 548--555.

\bibitem{moss2008box}
D.~Moss and P.~Levis, ``{BoX-MACs}: Exploiting physical and link layer
  boundaries in low-power networking,'' \emph{Computer Systems Laboratory
  Stanford University}, 2008.

\bibitem{sun2008ri}
Y.~Sun, O.~Gurewitz, and D.~B. Johnson, ``{RI-MAC}: A receiver-initiated
  asynchronous duty cycle {MAC} protocol for dynamic traffic loads in wireless
  sensor networks,'' in \emph{Proceedings of the 6th ACM Conference on Embedded
  Network Sensor Systems}.\hskip 1em plus 0.5em minus 0.4em\relax ACM, 2008,
  pp. 1--14.

\bibitem{alliance2007zigbee}
Z.~Alliance, ``Zigbee-2007 specification,'' \emph{Online:
  http://www.zigbee.org/Specifications/ZigBee/Overview.aspx}, 2007.

\bibitem{4215510}
T.~Avram, S.~Oh, and S.~Hariri, ``Analyzing attacks in wireless ad hoc network
  with self-organizing maps,'' in \emph{5th Annual Conference on Communication
  Networks and Services Research}, 2007, pp. 166--175.

\bibitem{de2002artificial}
L.~N. De~Castro and J.~Timmis, \emph{Artificial immune systems: A new
  computational intelligence approach}.\hskip 1em plus 0.5em minus 0.4em\relax
  Springer, 2002.

\bibitem{7845239}
G.~J. Pottie and A.~Pandya, \emph{Quality of service in wireless sensor
  networks}.\hskip 1em plus 0.5em minus 0.4em\relax John Wiley \& Sons, Inc.,
  2008, pp. 401--435.

\bibitem{chen2004qos}
D.~Chen and P.~K. Varshney, ``{QoS} support in wireless sensor networks: A
  survey,'' in \emph{International Conference on Wireless Networks}, vol. 233,
  2004.

\bibitem{Osborne:2008:TRI:1371607.1372727}
M.~A. Osborne, S.~J. Roberts, A.~Rogers, S.~D. Ramchurn, and N.~R. Jennings,
  ``Towards real-time information processing of sensor network data using
  computationally efficient multi-output {G}aussian processes,'' in
  \emph{Proceedings of the 7th International Conference on Information
  Processing in Sensor Networks}.\hskip 1em plus 0.5em minus 0.4em\relax IEEE
  Computer Society, 2008, pp. 109--120.

\bibitem{4230960}
N.~Ouferhat and A.~Mellouk, ``A {QoS} scheduler packets for wireless sensor
  networks,'' in \emph{International Conference on Computer Systems and
  Applications}, 2007, pp. 211--216.

\bibitem{4496881}
M.~Seah, C.-K. Tham, V.~Srinivasan, and A.~Xin, ``Achieving coverage through
  distributed reinforcement learning in wireless sensor networks,'' in
  \emph{3rd International Conference on Intelligent Sensors, Sensor Networks
  and Information}.\hskip 1em plus 0.5em minus 0.4em\relax IEEE, 2007, pp.
  425--430.

\bibitem{5284227}
R.~Hsu, C.-T. Liu, K.-C. Wang, and W.-M. Lee, ``{QoS}-aware power management
  for energy harvesting wireless sensor network utilizing reinforcement
  learning,'' in \emph{International Conference on Computational Science and
  Engineering}, vol.~2.\hskip 1em plus 0.5em minus 0.4em\relax IEEE, 2009, pp.
  537--542.

\bibitem{4976244}
X.~Liang, M.~Chen, Y.~Xiao, I.~Balasingham, and V.~C.~M. Leung, ``A novel
  cooperative communication protocol for {QoS} provisioning in wireless sensor
  networks,'' in \emph{5th International Conference on Testbeds and Research
  Infrastructures for the Development of Networks Communities and Workshops},
  2009, pp. 1--6.

\bibitem{Baccour:2012:RLQ:2240116.2240123}
N.~Baccour, A.~Koubaa, L.~Mottola, M.~A. Zuniga, H.~Youssef, C.~A. Boano, and
  M.~Alves, ``Radio link quality estimation in wireless sensor networks: A
  survey,'' \emph{ACM Transactions on Sensor Networks (TOSN)}, vol.~8, no.~4,
  p.~34, 2012.

\bibitem{Woo:2003:TUC:958491.958494}
A.~Woo, T.~Tong, and D.~Culler, ``Taming the underlying challenges of reliable
  multihop routing in sensor networks,'' in \emph{Proceedings of the 1st
  International Conference on Embedded Networked Sensor Systems}, ser. SenSys
  '03.\hskip 1em plus 0.5em minus 0.4em\relax New York, NY, USA: ACM, 2003, pp.
  14--27.

\bibitem{4428658}
K.~Shah and M.~Kumar, ``Distributed independent reinforcement learning ({DIRL})
  approach to resource management in wireless sensor networks,'' in
  \emph{Internatonal Conference on Mobile Adhoc and Sensor Systems}, 2007, pp.
  1--9.

\bibitem{Mainwaring:2002:WSN:570738.570751}
A.~Mainwaring, D.~Culler, J.~Polastre, R.~Szewczyk, and J.~Anderson, ``Wireless
  sensor networks for habitat monitoring,'' in \emph{Proceedings of the 1st ACM
  International Workshop on Wireless Sensor Networks and Applications}.\hskip
  1em plus 0.5em minus 0.4em\relax ACM, 2002, pp. 88--97.

\bibitem{Nadimi2008167}
E.~Nadimi, H.~T. S{\o}gaard, and T.~Bak, ``Zigbee-based wireless sensor
  networks for classifying the behaviour of a herd of animals using
  classification trees,'' \emph{Biosystems Engineering}, vol. 100, no.~2, pp.
  167--176, 2008.

\bibitem{4689144}
L.~Paladina, A.~Biundo, M.~Scarpa, and A.~Puliafito, ``Self organizing maps for
  synchronization in wireless sensor networks,'' in \emph{New Technologies,
  Mobility and Security}, 2008, pp. 1--6.

\bibitem{5196704}
O.~Postolache, J.~Pereira, and P.~Girao, ``Smart sensors network for air
  quality monitoring applications,'' \emph{IEEE Transactions on Instrumentation
  and Measurement}, vol.~58, no.~9, pp. 3253--3262, 2009.

\bibitem{gao2013wireless}
Y.~Gao, Y.~Lin, and Y.~Sun, ``A wireless sensor network based on the novel
  concept of an {I}-matrix to achieve high-precision lighting control,''
  \emph{Building and Environment}, vol.~70, pp. 223--231, 2013.

\bibitem{1425113}
N.~Kimura and S.~Latifi, ``A survey on data compression in wireless sensor
  networks,'' in \emph{International Conference on Information Technology:
  Coding and Computing}, vol.~2, 2005, pp. 8--13 Vol. 2.

\bibitem{Barr:2006:ELD:1151690.1151692}
K.~C. Barr and K.~Asanovi\'{c}, ``Energy-aware lossless data compression,''
  \emph{ACM Transactions on Computer Systems}, vol.~24, no.~3, pp. 250--291,
  Aug. 2006.

\bibitem{4472248}
J.~Haupt, W.~Bajwa, M.~Rabbat, and R.~Nowak, ``Compressed sensing for networked
  data,'' \emph{IEEE Signal Processing Magazine}, vol.~25, no.~2, pp. 92--101,
  2008.

\bibitem{5502565}
J.~Luo, L.~Xiang, and C.~Rosenberg, ``Does compressed sensing improve the
  throughput of wireless sensor networks?'' in \emph{International Conference
  on Communications}, 2010, pp. 1--6.

\bibitem{5707037}
S.~Feizi, M.~Medard, and M.~Effros, ``Compressive sensing over networks,'' in
  \emph{48th Annual Allerton Conference on Communication, Control, and
  Computing (Allerton)}, 2010, pp. 1129--1136.

\bibitem{7785419}
J.~B. Predd, S.~Kulkarni, and H.~V. Poor, \emph{Distributed learning in
  wireless sensor networks}.\hskip 1em plus 0.5em minus 0.4em\relax IEEE, 2006,
  vol.~23, no.~4, pp. 56--69.

\bibitem{Koby:4525469}
K.~Crammer, A.~Kulesza, and M.~Dredze, ``\BIBforeignlanguage{English}{Adaptive
  regularization of weight vectors},''
  \emph{\BIBforeignlanguage{English}{Machine Learning}}, vol.~91, no.~2, pp.
  155--187, 2013.

\bibitem{Yang:2009:OLE:1553374.1553521}
L.~Yang, R.~Jin, and J.~Ye, ``Online learning by ellipsoid method,'' in
  \emph{Proceedings of the 26th Annual International Conference on Machine
  Learning}, ser. ICML '09.\hskip 1em plus 0.5em minus 0.4em\relax New York,
  NY, USA: ACM, 2009, pp. 1153--1160.

\bibitem{wang2012exact}
J.~Wang, P.~Zhao, and S.~C. Hoi, ``Exact soft confidence-weighted learning,''
  in \emph{Proceedings of the 29th International Conference on Machine
  Learning}, 2012, pp. 121--128.

\bibitem{1413463}
J.~H. Kotecha, V.~Ramachandran, and A.~Sayeed, ``Distributed multitarget
  classification in wireless sensor networks,'' \emph{IEEE Journal on Selected
  Areas in Communications}, vol.~23, no.~4, pp. 703--713, 2005.

\bibitem{jiang2007architecture}
X.~Jiang, J.~Taneja, J.~Ortiz, A.~Tavakoli, P.~Dutta, J.~Jeong, D.~E. Culler,
  P.~Levis, S.~Shenker \emph{et~al.}, ``An architecture for energy management
  in wireless sensor networks,'' \emph{SIGBED Review}, vol.~4, no.~3, pp.
  31--36, 2007.

\bibitem{Zhang:1996:BED:235968.233324}
T.~Zhang, R.~Ramakrishnan, and M.~Livny, ``{BIRCH}: An efficient data
  clustering method for very large databases,'' \emph{ACM SIGMOD Record},
  vol.~25, no.~2, pp. 103--114, Jun. 1996.

\bibitem{Guha:1998:CEC:276304.276312}
S.~Guha, R.~Rastogi, and K.~Shim, ``{CURE}: An efficient clustering algorithm
  for large databases,'' in \emph{Proceedings of the 1998 ACM SIGMOD
  International Conference on Management of Data}, ser. SIGMOD '98.\hskip 1em
  plus 0.5em minus 0.4em\relax New York, NY, USA: ACM, 1998, pp. 73--84.

\end{thebibliography}

\end{document}